\documentclass[sigconf, screen, 10pt]{acmart}
\AtBeginDocument{%
  }

\usepackage{graphicx, url, xspace, xcolor, multirow, booktabs, amsfonts, subcaption, hyperref, colortbl, amsthm,  siunitx, enumitem}
\usepackage{algorithm}
\usepackage{algorithmic}




\newtheorem{problem}{Problem}



\makeatletter
\def\@ACM@copyright@check@cc{}
\makeatother 

\newcommand{\colorcell}[1]{
    \ifnum#1=100 \cellcolor{green!70!white}#1\% 
    \else
    \ifnum#1>80 \cellcolor{green!40!yellow}#1\% 
    \else
    \ifnum#1>60 \cellcolor{yellow!70!green}#1\% 
    \else
    \ifnum#1>40 \cellcolor{yellow!70!orange}#1\% 
    \else
    \ifnum#1>20 \cellcolor{orange!60!yellow}#1\% 
    \else
    \ifnum#1>0 \cellcolor{red!20!orange}#1\% 
    \else
    \cellcolor{red!100!white}#1\%  
    \fi
    \fi
    \fi
    \fi
    \fi
    \fi
}

\newcommand{\drex}{D-Rex}
\newcolumntype{L}{>{\raggedright\arraybackslash}X}

\newcommand{\blue}[1]{{\color{blue}\emph{#1}}}

\newif\ifdraft
\drafttrue 

\ifdraft
    \newcommand{\dante}[1]{{\color{purple} \textbf{Dante:} #1}}
    \newcommand{\sicheng}[1]{{\color{pink} \textbf{Sicheng:} #1}}
    \newcommand{\jesus}[1]{{\color{green} \textbf{Jesus:} #1}}
    \newcommand{\hai}[1]{{\color{magenta} \textbf{Hai:} #1}}
    \newcommand{\bogdan}[1]{{\color{violet} \textbf{Bogdan:} #1}}
    \newcommand{\greg}[1]{{\color{brown} \textbf{Greg:} #1}}
    \newcommand{\valerie}[1]{{\color{blue} \textbf{Valerie:} #1}}
    \newcommand{\haochen}[1]{{\color{teal} \textbf{Haochen:} #1}}
    \newcommand{\maxime}[1]{{\color{orange} \textbf{Maxime:} #1}}
    \newcommand{\kyle}[1]{{\color{red} \textbf{Kyle:} #1}}
    \newcommand{\ian}[1]{{\color{cyan} \textbf{Ian:} #1}}
\else
    \newcommand{\dante}[1]{}
    \newcommand{\sicheng}[1]{}
    \newcommand{\jesus}[1]{}
    \newcommand{\hai}[1]{}
    \newcommand{\bogdan}[1]{}
    \newcommand{\greg}[1]{}
    \newcommand{\valerie}[1]{}
    \newcommand{\haochen}[1]{}
    \newcommand{\maxime}[1]{}
    \newcommand{\kyle}[1]{}
    \newcommand{\ian}[1]{}
\fi

\author{Maxime Gonthier}
\affiliation{%
  \institution{University of Chicago}
  \institution{Argonne National Laboratory}
  \city{Chicago}
  \country{USA}}
\email{mgonthier@uchicago.edu}

\author{Dante D. Sanchez-Gallegos}
\affiliation{%
  \institution{Universidad Carlos III de Madrid}
  \city{Leganes}
  \country{Spain}}
\email{dantsanc@pa.uc3m.es}

\author{Haochen Pan}
\affiliation{%
  \institution{University of Chicago}
  \city{Chicago}
  \country{USA}}
\email{haochenpan@uchicago.edu}

\author{Bogdan Nicolae}
\affiliation{%
  \institution{Argonne National Laboratory}
  \city{Lemont}
  \country{USA}}
\email{bogdan.nicolae@acm.org}

\author{Sicheng Zhou}
\affiliation{%
  \institution{Southern University of Science and Technology}
  \city{Shenzhen}
  \country{China}}
\email{zhousc2021@mail.sustech.edu.cn}

\author{Hai Duc Nguyen}
\affiliation{%
  \institution{University of Chicago}
  \institution{Argonne National Laboratory}
  \city{Chicago}
  \country{USA}}
\email{ndhai@uchicago.edu}

\author{Valerie Hayot-Sasson}
\affiliation{%
  \institution{University of Chicago}
  \institution{Argonne National Laboratory}
  \city{Chicago}
  \country{USA}}
\email{vhayot@uchicago.edu}

\author{J. Gregory Pauloski}
\affiliation{%
  \institution{University of Chicago}
  \city{Chicago}
  \country{USA}}
\email{jgpauloski@uchicago.edu}

\author{Jesus Carretero}
\affiliation{%
  \institution{Universidad Carlos III de Madrid}
  \city{Leganes}
  \country{Spain}}
\email{jcarrete@inf.uc3m.es}

\author{Kyle Chard}
\affiliation{%
  \institution{University of Chicago}
  \institution{Argonne National Laboratory}
  \city{Chicago}
  \country{USA}}
\email{chard@uchicago.edu}

\author{Ian Foster}
\affiliation{%
  \institution{University of Chicago}
  \institution{Argonne National Laboratory}
  \city{Chicago}
  \country{USA}}
\email{foster@anl.gov}

\renewcommand{\shortauthors}{Gonthier, et al.}

\title{\drex{}: Heterogeneity-Aware Reliability Framework and Adaptive Algorithms for Distributed Storage}

\acmConference[ICS '25]{2025 International Conference on Supercomputing}{June 8--11, 2025}{Salt Lake City, UT, USA}
\acmDOI{10.1145/3721145.3730412}
\acmISBN{979-8-4007-1537-2/2025/06}

\begin{document}

\begin{abstract}
The exponential growth of data necessitates distributed storage models, such as peer-to-peer systems and data federations.
While distributed storage can reduce costs and increase reliability, the heterogeneity in storage capacity, I/O performance, and failure rates of storage resources makes their efficient use a challenge.
Further, node failures are common and can lead to data unavailability and even data loss. 
Erasure coding is a common resiliency strategy implemented in storage systems to mitigate failures by striping data across storage locations.
However, erasure coding is computationally expensive and existing systems do not consider the heterogeneous resources and their varied capacity and performance when placing data chunks. 
We tackle the challenges of using erasure coding with distributed and heterogeneous nodes, aiming to store as much data as possible, minimize encoding and decoding time, and meeting user-defined reliability requirements for each data item.
We propose two new dynamic scheduling algorithms, D-Rex LB and D-Rex SC, that adaptively choose erasure coding parameters and map chunks to heterogeneous nodes.
D-Rex SC achieves robust performance for both storage utilization and throughput, at a higher computational cost, while D-Rex LB is faster but with slightly less competitive performance.
In addition, we propose two greedy algorithms, GreedyMinStorage and GreedyLeastUsed, that optimize for storage utilization and load balancing, respectively.
Our experimental evaluation shows that our dynamic schedulers store, on average, 45\% more data items without significantly degrading I/O throughput compared to state-of-the-art algorithms, while GreedyLeastUsed is able to store 21\% more data items while also increasing throughput.
\end{abstract}



\maketitle

\section{Introduction}\label{sec:intro}

The increasing size of data, coupled with its distributed production and acquisition, is driving the adoption of distributed data storage models. Distributed storage is often seen as a way of reducing costs, increasing fault tolerance, and improving performance. However, achieving these benefits requires addressing the challenges posed by heterogeneous storage nodes (potentially geographically distributed) with varying capacities, bandwidths, and failure rates. 

A key challenge in this context is the resilient storage of data in a cost-efficient manner. Erasure coding (EC)~\cite{7823943} is a well-known technique that provides the same resilience guarantees as full data replication but uses significantly less storage space  
at the cost of additional computations.
EC techniques divide a data item $d$ into $K$ chunks of size $size(d)/K$ and add $P$ parity chunks of the same size, such that any $K$ of the $K+P$ chunks are sufficient to recover the original item through a decoding operation. Thus, if each of the $K+P$ chunks is placed on a different storage system, up to $P$ failures can be tolerated before data items are lost.

\begin{figure}
    \centering
    \includegraphics[width=\linewidth]{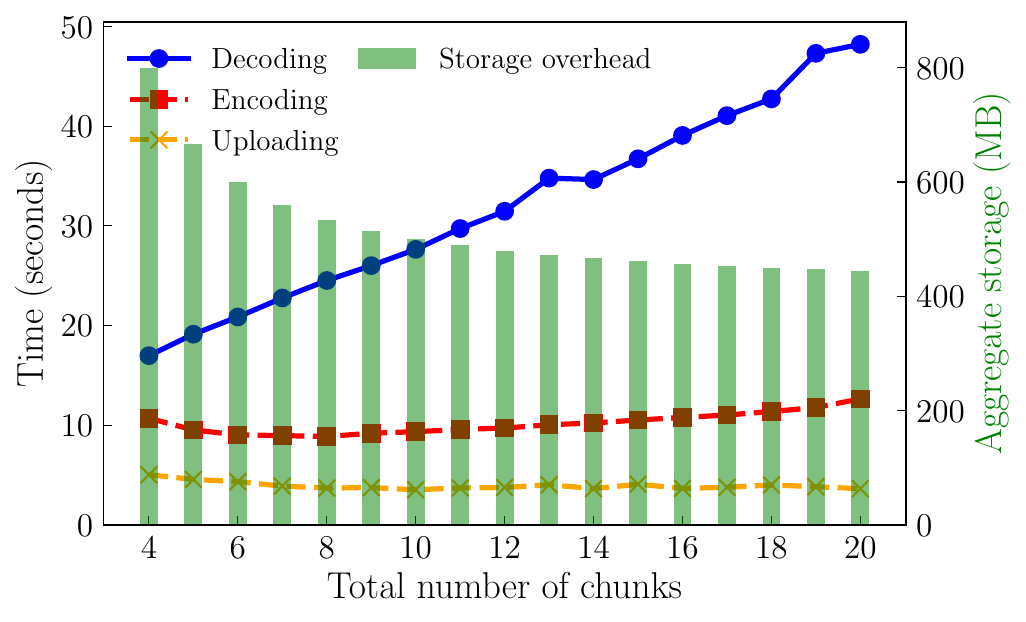}
    \caption{Breakdown of encode, decode, and upload times to store  400~MB across a varying number of nodes based on~\cite{9180305} on a 48-core Intel Xeon E5-2670. The parity chunks $P=2$, resulting in $K$ data chunks equal to total number of chunks minus two.
    }
    \Description{Aggregate storage cost and encode, decode, and upload times to store a 400~MB data item across a varying number of nodes using a similar encode and decode process.}    
    \label{fig:EC_time}
    \vspace{-10pt}
\end{figure}

However, EC faces two main challenges.
First, it incurs costs due to the complexity of chunking objects and calculating the parity data~\cite{zebra, 9180305}: costs that tend to increase with the number of chunks.
To illustrate these costs, we conducted an experiment in which we encoded and decoded a 400 MB data item with a fixed $P=2$ and varying values of $K$.
As shown in \autoref{fig:EC_time}, increased $K$ reduces storage overhead but increases decoding time.
Second, state-of-the-art distributed storage algorithms employ EC using fixed values of $K$ and $P$ and distribute chunks uniformly,
under the assumption that all data items are equal and the storage
nodes are homogeneous. Given the diversity of storage requirements
(e.g., some data items require a higher reliability guarantee and/or accessed faster than others) and heterogeneity of storage nodes (e.g., due to federation of storage systems), this assumption is outdated and necessitates more adaptive EC.

To address these challenges, we propose \drex{} (Dynamic Resilience Extension), an adaptive EC algorithm based on an innovative reliability model that allows a customized reliability target for each stored data item. Based on this reliability model, we contribute two greedy algorithms and two adaptive algorithms that automatically decide the number of data chunks, parity chunks, and allocation of chunks on storage nodes in order to satisfy reliability targets, maximize storage space utilization, and minimize the I/O overhead of data access.
\drex{} is grounded in real-world applications and can be integrated into existing data transfer and storage infrastructures, such as Globus~\cite{globuspaas} or DynoStore~\cite{dynostore,sanchez2025dynostore}.
We summarize our contributions as follows:
\begin{enumerate}[topsep=0pt,itemsep=0pt,leftmargin=12pt]
    \item We introduce an innovative reliability model that leverages the notion of a \emph{reliability target}, expressed as the chance of successful access to a data item within a given time interval, making it easy for users to express and reason about quality-of-service guarantees (\S\ref{sec:reliablity_model}).
    \item We introduce two greedy algorithms, GreedyMinStorage and GreedyLeastUsed, and two dynamic algorithms, \drex{}~LB and \drex{}~SC, based on our reliability model. The \drex{} algorithms aim to solve the multi-objective optimization problem (\S\ref{sec:formulation}) of successfully storing as much data as possible, satisfying the reliability target and storage capacity constraints, and minimizing I/O overheads associated with read/write operations (\S\ref{sec:algos}).
    \item We design and implement a dynamic data storage simulator that we employ to evaluate our proposed algorithms against the state-of-the-art algorithms (\S\ref{sec:sota}) under controlled settings. Using the simulator, we run extensive simulations using diverse datasets, showing that \drex{}~SC and \drex{}~LB store on average 45\% and 31\% more data while only 0.4 and 0.3 MB/s slower, respectively, than state-of-the-art algorithms (\S\ref{sec:expes}) when using heterogeneous nodes.
    GreedyLeastUsed increases the amount of data stored by up to 21\% while also increasing throughput.
    \item We integrate our algorithms into a real memory system to perform experiments in an unconstrained memory scenario and show that our algorithms do not significantly degrade throughput under such conditions (\S\ref{sec:real}).
\end{enumerate}

\section{Related Work}
\label{sec:relatedwork}

\textbf{Distributed Storage:}
Distributed object stores and file systems often rely on replication, where full copies of a data item are stored on different nodes to ensure availability.
AWS S3~\cite{aws-s3} replicates objects across at least three availability zones.
Similarly, Hadoop Distributed File System  (HDFS) replicates each data item twice~\cite{hard}, resulting in a total of three copies and a storage overhead of 200\%.
The Inter Planetary File System~\cite{doan2022toward} is a peer-to-peer (P2P) file system that fully replicates data items as they are shared between different peers and, by the nature of P2P, must manage many heterogeneous nodes.
SkyStore~\cite{liu2025skystore} dynamically replicates and evicts objects across different cloud storage providers according to application access patterns with the goal of optimizing for lower cost.
Our solutions differ by using erasure coding, which has lower storage overhead than full replication.

\textbf{Erasure Coding in Distributed Storage:}
HDFS~\cite{hdfsdocrs} uses the well-established Reed-Solomon~\cite{wicker1999reed} erasure coding scheme and stores data and parity chunks in a distributed fashion to minimize inter-rack write traffic. It relies on a centralized NameNode to manage the encoding process.
Gluster~\cite{glusterfsredhat} is a distributed file system that combines disk storage from multiple servers into a single namespace. Gluster integrates erasure coding with a default configuration of 4 data chunks and 2 parity chunks.
It uses dispersed bricks rather than dedicated nodes to store chunks, allowing for more granular storage.
Distributed Asynchronous Object Storage (DAOS) is an object store where objects are managed as collections, called containers, that can scale across multiple storage devices~\cite{10.1145/3581576.3581577}.
It implements EC and allows users to define the desired level of reliability.
The storage efficiency of EC has motivated its implementation in distributed data storage systems such as EdgeHydra~\cite{he2024edgehydra} and ELECT~\cite{ren2024elect}. In these systems, erasure codes stripe data items across multiple storage nodes to balance workloads and enable parallel data access for improved performance.
OceanStore~\cite{kubiatowicz2000oceanstore} and ER-Store ~\cite{li2021er} implement a combination of EC and replication to provide high availability of data, assuming that all nodes are prone to fail.
While these systems efficiently manage EC, they assume that all nodes are homogeneous, and as a result, the number of data and parity chunks can be static. Our approach differs by incorporating heterogeneity into decisions on how to utilize EC and where to place chunks.




\textbf{Erasure Coding Optimizations:}
SMFRepair~\cite{zhou2022bandwidth} suggests saturating the bandwidth of the least busy node and transferring data from links with the largest possible bandwidth to minimize the transfer time when decoding data. We consider a simpler model where we do not include transfers between storage nodes.
\cite{cost_effective_edge} propose a linear programming model for the data block placement problem at the edge in order to minimize storage cost while maintaining quality of service.
Some approaches use TOPSIS to optimize chunk placement on heterogeneous storage nodes to improve read/write throughput~\cite{topsis_ceph_hetero} but do not consider storage size limitations.
While these solutions optimize for either transfer time or storage cost, we have developed an approach that optimizes for both time and space in the context of data storage on heterogeneous nodes with limited storage.


\textbf{Failure‐Recovery Techniques}
Erasure coding (EC) significantly lowers repair bandwidth compared to full replication~\cite{10.1007/3-540-45748-8_31}.
To further reduce recovery latency, schemes such as proactive repair rebuild lost chunks immediately upon failure detection, or even preemptively when failures are predicted~\cite{9721141,10.1145/3708994}. These techniques focus solely on the repair phase, and can be layered on top of our chunking and placement model without any modifications. Checkpoint-restart is another actively researched area. Several efforts leverage multiple storage levels to asynchronously hide the I/O overhead of persisting checkpoints~\cite{VeloCIPDPS19, VELOC-SuperCheck21, VELOC-FGCS24}. EC is used to protect the node-local checkpoints against node failures, as an alternative to storing checkpoints to a parallel file system. 

\section{Framework}\label{sec:model}

Here, we formalize the problem statement. First, we introduce an innovative reliability model that addresses the question of satisfying resilience based on user-specified quality of service constraints. Then, we 
formulate an optimization problem to decide how to encode and distribute multiple data items that need to be written to a resilient data repository subject to our reliability model. Key notations are summarized in \autoref{tab:notations}.

\begin{table}[t]
\centering
\caption{List of notations.}
\scalebox{0.84}{
\begin{tabular}{>{\bfseries}l p{5cm}}
    \toprule
    \textbf{Notation} & \textbf{Description} \\
    \midrule

    \multicolumn{2}{l}{\textbf{Known to the algorithms}} \\
    \midrule
    $L$ & Number of available storage nodes. \\
    $\mathbb{S} = \{S_1, \ldots, S_L\}$ & The set of storage nodes. \\
    $size(S_i)$ & Total storage size of node $S_i$. \\
    $F(S_i, t)$ & Free storage size on node $S_i$ at time $t$. 
    \\
    $Pr_{\textrm{failure}}(S_i, \Delta t)$ & Probability of node $S_i$ failing at least once over a time period $\Delta t$.\\
    $B_r(S_i)$ & Read bandwidth of $S_i$, in MB/s.\\
    $B_w(S_i)$ & Write bandwidth of $S_i$, in MB/s.\\
    $T_{\textrm{encode}}(N_d,K_d,size(d))$ & Time required to encode a data item $d$ of size $size(d)$ into $N_d$ chunks, each of size $size(d)/K_d$.\\
    $T_{\textrm{decode}}(K_d,size(d))$ & Time required to recreate a data item of size $size(d)$ from $K_d$ chunks.\\
    
    \midrule
    \multicolumn{2}{l}{\textbf{Known for each data item $d$ at the time that it is to be stored}} \\
    \midrule
    $size(d)$ & Size of $d$ in MB.\\
	  $\Delta t_d$ & Duration $d$ has to stay available.\\
    $\mathit{RT(d)}$ & Reliability target of $d$.\\
    
    \midrule
    \multicolumn{2}{l}{\textbf{Unknown to the algorithms}} \\
    \midrule
    $m$ & Number of data items to store. \\
    $\mathbb{D} = \{d_1, \ldots, d_{m}\}$ & The set of data items to store. \\
    
    \midrule
    \multicolumn{2}{l}{\textbf{Chosen by the algorithms for each data item $d$}} \\
    \midrule
	$K_d$ & Number of data chunks required to retrieve data item. Controls the data item's chunk size ($=\lceil size(d)/K_d\rceil$).\\
    $P_d$ & Number of parity chunks added for $d$. \\
    $\mathcal{C}_d$ & Set of chunks for data item $d$. \\
    $N_d$ & Total number of chunks: $N_d=K_d+P_d$. \\
    $\mathcal{M}_d$ & Subset of nodes in $\mathbb{S}$ where the chunks of $\mathcal{C}_d$ have been loaded. $|\mathcal{C}_d| = |\mathcal{M}_d| = K_d + P_d$ \\
    \bottomrule
    \end{tabular}
}
\label{tab:notations}
\end{table}

\subsection{Reliability Model}
\label{sec:reliablity_model}

As a first contribution, we introduce a reliability model that allows the user to input the level of reliability they want for their data. The goal is to give the user as much flexibility as possible without over-complicating how the user must reason about resilience. Intuitively, each data item that needs to be stored has a well-defined life cycle. For example, the intermediate results of an application need to persist for the lifetime of the application. Experimental results from scientific applications typically need to be persistent for days or months until data are analyzed and summarized. Some important results need to be archived and retained for years. Thus, each data item $d$ (out of a total of $m$ data items) to be stored in the repository has a retention time $\Delta t_d<\infty$ (i.e., an expiration). We propose the notion of a reliability target $RT(d)$, a value between $(0,1)$ representing the chance that $d$ can be successfully retrieved before expiration. The user only specifies $\Delta t_d$ and $RT(d)$.

\begin{figure}
\includegraphics[width=1.01\linewidth, trim={2em 1em 2em 1em}, clip]{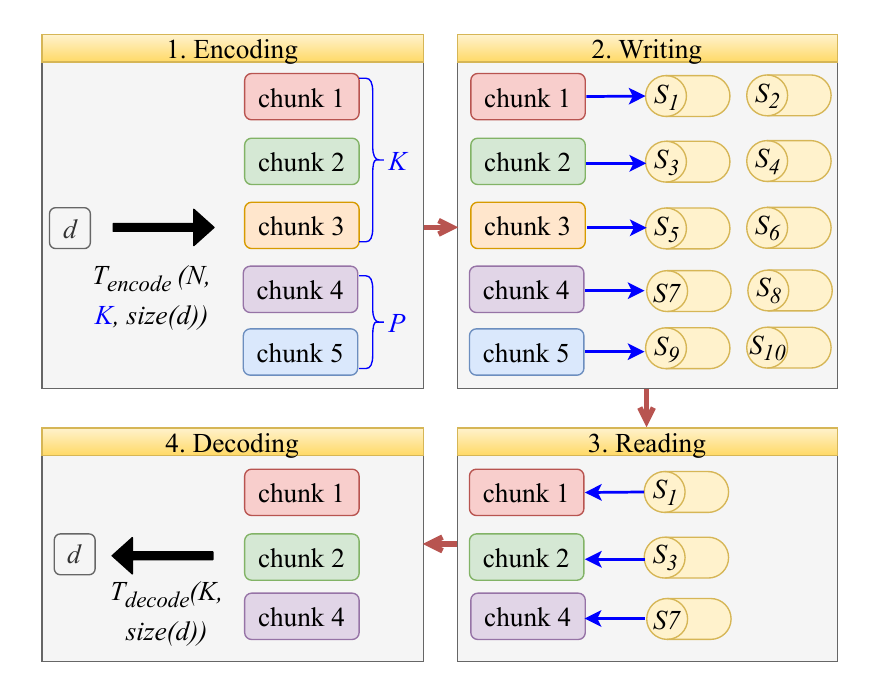}
\caption{
Data storage and retrieval, from top left, clockwise, with algorithm decisions in blue.
(1) \textbf{Encoding} of data $d$ into $N$ = \blue{K} data chunks + \blue{P} parity chunks;
(2) \textbf{Writing} of chunks to storage nodes, with choices shown by {\color{blue}$\rightarrow$};
(3) \textbf{Reading} of $K$ chunks from selected nodes (more choices, {\color{blue}$\leftarrow$}); and
(4) \textbf{Decoding} of original content from read chunks.}
\label{fig:model}
\Description{Data storage and retrieval, from top left, clockwise, with algorithm decisions in blue.}
\end{figure}

A repository consists of $L$ storage nodes $\mathbb{S} = \{S_1, \ldots, S_L\}$. A storage node may fail at any time. We assume an erasure-coded representation that splits each data item $d$ into a chunking $\mathcal{C}_d$ of $K_d$ equally sized data chunks, complemented by $P_d$ parity chunks of the same size. Each chunk is stored on a different node. We denote the mapping of chunks to nodes as $\mathcal{M}_d$. If at least $K_d$ chunks of the total $K_d+P_d$ chunks are still accessible on nodes before $d$ expires, the data item is available. For example, \autoref{fig:model} shows the encoding of a data item into 3 data chunks and 2 parity chunks that are written to 5 distinct nodes with mapping $(S_1,S_3,S_5,S_7,S_9)$. Assuming $S_5$ has failed, the data item can be reconstructed from the two remaining data blocks and any parity block using a decoding operation. Encoding and decoding involve overheads $T_{encode}$ and $T_{decode}$, respectively.
Note that the erasure-coded representation is the most general one. For example, replication can be seen as a particular case in which $K_d=1$ represents the original data item, $P_d$ is the number of replicas, and $T_{encode}$ = $T_{decode} = 0$.

We assume a fail-stop failure model in which the storage nodes are either operational or have permanently failed.
Failure results in the permanent loss of any chunks stored on the node.
For simplicity, the probability of node failure is equivalent to the probability of disk failure (detailed in \S\ref{sec:backblaze_nodes}).
We assume the remaining components in the node will not fail, but this model can be extended to account for such failures.
Lastly, we assume that the failure rate of a storage node is constant. Then, the probability of a storage node failing at least once during a period $\Delta t$ (expressed as a fraction of a year):

\begin{equation}
\begin{split}
Pr_{\textrm{failure}}(S_i,\Delta t) = 1 - e^{-\lambda_{\text{rate}} \cdot \Delta t}
\label{eq:failure}
\end{split}
\end{equation}

Assuming failures follow a homogeneous Poisson process with a constant rate of occurrence $\lambda_{rate}$~\cite{elerath2008highly},  we can compute the probability of a data item $d$ with $P_d$ parity chunks and a mapping $\mathcal{M}_d$ to remain available for $\Delta t_d$ days: $Pr_{\textrm{avail}}(\mathcal{M}_d,P_d,\Delta t_d)$.
This is equivalent to the probability of no more than $P_d$ nodes failing. To this end, we use the cumulative Poisson binomial distribution function to calculate the probability of a certain number $X$ of events (i.e., node failures in our case) across a set of independent Bernoulli trials with probability $Pr_{\textrm{failure}}(S_d,\Delta t_d)$:
{
\begin{align}
& Pr_{\textrm{avail}}(\mathcal{M}_d,P_d,\Delta t_d) = Pr(X \leq P_d) = \notag \\
& \hspace{1ex} \sum _{l=0}^{P_d}\sum \limits _{A\in F_{l}}\prod \limits _{i\in A}Pr_{\textrm{failure}}(S_i,\Delta t_d)\prod \limits _{j\in A^{c}}\left(1-Pr_{\textrm{failure}}(S_j,\Delta t_d)\right)
\label{eq2}
\end{align}
}
with $F_l$ the set of all possible subsets of nodes in $\mathcal{M}_d$ containing exactly $l$ nodes.
\autoref{eq2} is precise but quadratic in complexity, so we implemented an approximation inspired by previous work \cite{HONG201341,poisson_python}. 
This approximation yields the \textit{reliability constraint} of each data item, defined as:
\begin{equation}
    Pr_{\textrm{avail}}(\mathcal{M}_d,P_d,\Delta t_d) \geq RT(d)
    \label{reliability-constraint}
\end{equation}
\autoref{reliability-constraint} is a key optimization constraint based on a simple user-specified reliability target.

While we assume uniform and independent node failures, our framework extends to any failure‐prediction model (e.g., as in~\cite{9671605}). We would only need to replace $Pr_{\textrm{failure}}(S_i,\Delta t)$ with the predicted probability of node $S_i$ to fail, given its current age, within the next $\Delta t$.
\autoref{eq2} could then be applied without any modification.

\subsection{Problem Formulation}
\label{sec:formulation}

We assume the $m$ requests to store data items are ordered.
Multiple users may submit requests concurrently, but for the purposes of space allocation, each item must be considered sequentially.
This does not preclude physically writing many items concurrently. Each data item in $\mathbb{D} = \{d_1, \ldots, d_{m}\}$ is associated with a submission timestamp $t_i > t_j$ for each $i > j$, repository-assigned based on concurrency control policies.

For each storage node, the following information is available: its total capacity, $size(S_i)$; free space at time $t$, $F(S_i, t)$; write bandwidth, $B_w(S_i)$; read bandwidth, $B_r(S_i)$; and probability of failure over a given $\Delta t$, $Pr_{\textrm{failure}}(S_i,\Delta t)$: see \autoref{eq:failure}. Our goal is to automatically determine for each data item $d$: (i) the number of data chunks, $K_d$; (ii) the number of parity chunks, $P_d$; and (iii) the mapping $\mathcal{M}_d$ of these chunks onto a subset of $K_d+P_d$ storage nodes in $\mathbb{S}$, while optimizing two quality metrics: (i) the total size of data \textbf{items successfully stored} and (ii) the \textbf{I/O overhead} of the write/read operations. Furthermore, the optimization must be done in an online fashion, meaning any decision about $d_i$ can only depend on decisions taken for previous data items $d_j$ with $j < i$, i.e., foreknowledge of storage requests is not available.

The motivation behind the quality metrics is based on two observations: (i) the goal of a storage repository
is to maximize effective use of its capacity and (ii) users desire requests to be served quickly in addition to successfully. 
These two quality metrics may require a trade-off: increasing the success rate of write requests may come at the cost of greater I/O overhead.
We do not prioritize one metric because both are important.

More formally, we consider a write operation successful if and only if it simultaneously satisfies two constraints: (i) there is capacity available on each node in $\mathcal{M}_d$ to store the new chunk and (ii) the reliability constraint $Pr_{\textrm{avail}}(\mathcal{M}_d,P_d,\\\Delta t_d) \ge RT(d)$ (introduced in \autoref{reliability-constraint}) can be satisfied. Then, the total amount of data successfully stored is defined as:

\[
\mathbb{W} = \sum_{d \in \mathcal{D}_{\text{success}}} size(d)
\]
where:
\[
\mathcal{D}_{\textrm{success}} = d \in \mathbb{D} \mid
\begin{aligned}
& Pr_{\textrm{avail}}(\mathcal{M}_d, P_d, \Delta t_d) \geq RT(d) \\
& \textrm{and} \ \forall S_i \in \mathcal{M}_d, \ F(S_i, t_d) \geq \frac{size(d)}{K_d}
\end{aligned}
\]

We are concerned with I/O overheads $\mathbb{T}$ required to encode, decode, write, and read each successfully stored data item.
These overheads depend on $K_d$, $P_d$ and $\mathcal{M}_d$.
The relative importance of each overhead can be weighed depending on the application and user requirements, but we assume overheads share equal importance for the purposes of this work.
Thus, we define average I/O throughput as $\mathbb{T}$ where:

\[
\mathbb{T} = \frac{\mathbb{W}}{\sum_{d \in \mathcal{D}_{\textrm{success}}} T_{\textrm{encode}} + T_{\textrm{decode}} + T_{\textrm{write}} + T_{\textrm{read}}}
\]

We assume that all chunk transfers are parallelized and that no two nodes share a connection, which would introduce contention.
Thus, the slowest node in $\mathcal{M}_d$ is the bottleneck for both writing ($T_{\text{write}}$) and reading ($T_{\text{read}}$).
We assume that encoding and decoding times are constant for a given configuration $K_d$, $P_d$, and $size(d)$.

Given the formal definitions of $\mathbb{W}$ and $\mathbb{T}$, we summarize the optimization problem in \autoref{pb:pb_def}.
\begin{problem}\label{pb:pb_def}
    Given a sequence of $m$ data items $\mathbb{D} = \{d_1, \ldots, \\d_m\}$ and $L$ storage nodes $\mathbb{S} = \{S_1, \ldots, S_L\}$, determine a \textbf{chunking}
    $\mathcal{C}_d$ of each data item $d \in \mathbb{D}$ as a set of $K_d$ data blocks and $P_d$ parity blocks, plus a \textbf{mapping} $\mathcal{M}_d$ that assigns each data block and parity block to a storage node, such as to maximize the average I/O throughput $\mathbb{T}$ and the total amount $\mathbb{W}$ of successfully stored data items while subject to capacity availability on each storage
    node in $\mathcal{M}_d$ to store the chunks in $\mathcal{C}_d$ and a target reliability constraint defined as $\mathcal{M}_d$ satisfy $Pr_{\textrm{avail}}(\mathcal{M}_d,P_d,\Delta t_d) \geq RT(d)$.
\end{problem}



\section{Dynamic Resilience Extension (\drex{})}\label{sec:algos}

To address the limitations described in \S\ref{sec:relatedwork}, in this section we present four algorithms that aim to solve the problem described in \S\ref{sec:formulation}. 
The first two algorithms we propose are greedy algorithms (\S\ref{sec:greedy} and \S\ref{sec:greedyleastused}). 
To address the limitations of these greedy algorithms, we propose two more algorithms. 
The first algorithm, presented in \S\ref{sec:drexlb}, aims to achieve good load balancing while having low complexity, making it easily applicable. The second algorithm, presented in \S\ref{sec:drexsc}, is more computationally expensive but seeks to optimize the trade-off between throughput and amount of data stored.
For each algorithm, we describe below the decision made when a new data item $d$ is stored at time $t$.

\subsection{Greedy Storage Minimization}\label{sec:greedy}

We assume that a reasonable solution for a cost minimization problem like Problem~\ref{pb:pb_def} is to minimize the total storage overhead for each data item.
We denote this algorithm \textit{GreedyMinStorage}.
It selects values for $N_{d}$ and $K_{d}$ that minimize the storage overhead for each data item
by solving an optimization problem such as:
\begin{equation}
    \begin{aligned}
         \text{minimize }\quad & \frac{size(d)}{K_{d}} \times N_{d}\\
        \text{s.t.}\quad  & Pr_{\textrm{avail}}(\mathcal{M}_{d},K_{d},\Delta t_{d}) \geq RT(d)
    \end{aligned}
\end{equation}
and use $P_{d}=N_{d}-K_{d}$ for the number of parity chunks.
The mapping $\mathcal{M}_{d}$ is chosen by favoring the nodes with the fastest bandwidth.
This usually corresponds to selecting high values of $N$ and $K$, so as to make each chunk as small as possible.
The intuition is that, first, smaller chunks are faster to transfer, so in the multi-objective problem we defined, this would ensure read/write efficiency.
Second, it incurs minimal storage overhead, allowing for more data items to be stored, thus satisfying the storage objective we defined.

\subsection{Greedy Maximum Free Space}\label{sec:greedyleastused}

The \textit{GreedyLeastUsed} algorithm prioritizes nodes with the most available storage when placing chunks.
This approach ensures that data is distributed to the least utilized nodes, balancing storage consumption across the system, which we believe is a reasonable approach that a system manager would want to use.
The algorithm selects values for $P_{d}$ and $K_{d}$ that are as small as possible in order to greedily reduce the encoding/decoding overhead:

\begin{equation}
\begin{aligned}
\text{minimize}\quad & P_d + K_d \\
\text{s.t.}\quad & Pr_{\textrm{avail}}(\mathcal{M}_d, K_d, \Delta t_d) \geq RT(d), \\
\text{and}\quad & \mathcal{M}_d = K_d + P_d \text{ nodes with highest } F(S_i)
\end{aligned}
\end{equation}

\subsection{\drex{} LB: Adding Load Balancing}\label{sec:drexlb}

The greedy algorithms presented above do not account for varying storage sizes.
The intuition behind \drex{} LB is to calculate a balance penalty for all nodes, regardless of whether a node stores a given item.
By assigning penalties to nodes that do not store the data item, the algorithm promotes a more even distribution of load across the system, enhancing load balancing and overall system efficiency.

The first step of the algorithm, detailed in \autoref{algo:drex-lb}, is to compute the average free capacity ($F_{\textrm{average}}$) of the storage nodes.
We then iterate over the number of data chunks $K_{d}$ required to reconstruct the original data item.
For nodes that would store the data item $d$, we calculate a balance penalty based on the node's free storage, the average free storage across all nodes, and the chunk size.
For nodes that would not store $d$, we still compute the penalty but exclude the chunk size.
When the average free capacity is high, indicating that most nodes are relatively empty, the chunk size has minimal impact on the balance penalty, and the algorithm prioritizes filling the largest nodes first.
Conversely, when the average free storage is small, the nodes begin to saturate and the chunk size has a greater impact on the balance penalty.
This incentives the algorithm to use larger values of $K_{d}$.
Ultimately, we choose the value of $K_{d}$ that yields the smallest penalty while also minimizing $P_{d}$ and ensuring the reliability target is met.

\drex{} LB requires looping $\frac{L(L+1)}{2}$ times over each mapping.
Verifying whether a mapping satisfies the reliability constraint requires computing the CDF for the Poisson-Binomial distribution, which in our implementation has a cost of at most $O(K^{2})$.
Since $K$ can be at most $L-1$, the worst-case complexity simplifies to $O(L^4)$. However, in most cases $K$ is small and the break condition at Line \ref{line:break} is reached early. Consequently, the scheduling overhead per data item is negligible compared to the data transfer time, as summarized in \autoref{tab:complexity}.

\begin{algorithm}[t]
    \caption{\drex{} Load Balancing (\drex{} LB)}\label{algo:drex-lb}
    \begin{algorithmic}[1]
        \STATE $F_{\textrm{average}} \gets \frac{1}{L} \sum\limits_{i=1}^{L} F(S_i, t_{d})$
        \STATE Sort $\mathbb{S}$ in order of decreasing $F(S_i, t_{d})$
        \STATE $min\_bp \gets \infty$ \hfill \textit{// bp: balance penalty}
        \STATE $min\_K \gets -1$
        \FOR{$P_{d} \gets 1$ to $P_{d} < L$}
            \FOR{$K_{d} \gets 2$ to $K_{d} \leq L - P_{d}$}
                \STATE $bp \gets 0$
                \STATE $\mathcal{M}_{d} \gets$ first $K_{d}+P_{d}$ nodes in sorted $\mathbb{S}$
                \IF{$Pr_{\textrm{avail}}(\mathcal{M}_{d},P_{d},\Delta t_{d}) \ge RT(d)$}
                    \FOR{each $S_j \in \mathcal{M}_{d}$}
                        \STATE $bp \mathrel{+}= \left|F(S_j, t_{d}) - size(d)/K_{d} - F_{\textrm{average}}\right|$
                    \ENDFOR
                    \FOR{each $S_j \notin \mathcal{M}_{d}$}
                        \STATE $bp \mathrel{+}= \left|F(S_j, t_{d}) - F_{\textrm{average}}\right|$
                    \ENDFOR
                    \IF{$bp < min\_bp$}
                        \STATE $min\_K \gets K_{d}$
                        \STATE $min\_bp \gets bp$
                    \ENDIF
                \ENDIF
            \ENDFOR
            \IF{$min\_K \neq -1$}
                \STATE Break \label{line:break}
            \ENDIF
        \ENDFOR
        \STATE Create  $min\_K + P_{d}$ chunks of size $size(d)/min\_K$ and store them on the first $min\_K + P_{d}$ nodes of sorted $\mathbb{S}$
    \end{algorithmic}
\end{algorithm}

\subsection{\drex{} SC: Considering System Capacity}\label{sec:drexsc}

We propose a second dynamic algorithm, \textit{\drex{} SC} (see \autoref{algo:drex-sc}), that is more computationally intensive than \drex{} LB but adapts better to node storage saturation.
The intuition is to consider three key parameters: (i) the cost of encoding and decoding, (ii) the erasure coding storage overhead for a given $N$ and $K$, and (iii) the proportion of free storage remaining across the nodes.

We begin by considering different node mappings (e.g., [1, 2, 3], [1, 2, 4], [1, 2, 3, 4], etc.) as described in Lines~\ref{line:mapping}--\ref{line:progress_saturation}.
We consider the first $2^{10}$ mappings of nodes with the highest available storage, starting with the top nodes sequentially. As an example, initially, we examine mappings of nodes [1, 2], then expand to [1, 2, 3] and so on up to [1, 2, ..., L]. Then, we explore other combinations such as [2, 3] etc.
Limiting to $2^{10}$ mappings ensures a balance between complexity and effective results at scale.

For each mapping $\mathcal{M}$, we know the number of nodes, $N = |\mathcal{M}|$, and the failure rate of each node in $\mathcal{M}$.
We then calculate the couple $K$, $P$ that minimize storage overhead while meeting the data reliability goal as well as the three key parameters: duration, storage cost, and saturation cost.

\begin{figure}
    \centering
    \includegraphics[width=\linewidth]{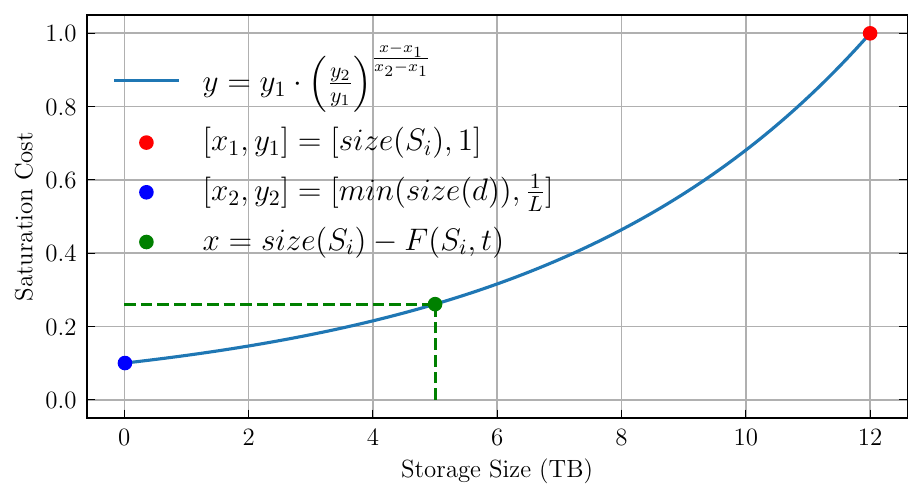}
    \caption{Example of the function used to calculate the saturation cost. The exponential curve spans from the smallest known data item size to the total storage capacity of the node. The green dot is an example of a node's available free storage.}
    \Description{Example of the function used to calculate the saturation cost.}
    \label{fig:exponential_system_saturation}
\end{figure}

\textit{Duration} comprises the encoding and decoding times and the parallelized read and write times (assumed to be the time for the node with the worst bandwidth) using the nodes in $\mathcal{M}$.
Linear regression is used to estimate the encode and decode times given a set of encode and decode times from different data item sizes and numbers of data and parity chunks.
In practice, our tests over data sizes ranging from 1 to 500 MB show that the linear regression prediction closely matches the actual measurements.
\textit{Storage cost} is the storage overhead of erasure coding for $N$ nodes with $K$ data chunks.
\textit{Saturation cost} aims to evaluate how close a node is to its storage limit. We assign a saturation score ranging from 0 to 1 based on the node's utilization relative to its maximum capacity. The scoring function (denoted as $f(x)$ on line \ref{line:progress_saturation}) follows an exponential curve to penalize nodes approaching their limit (see \autoref{fig:exponential_system_saturation}).
Completely filling a node is often a bad solution, as it reduces the number of nodes that can be used for future data items, thus forcing the use of smaller values of $N$, which is worse overall for storage overhead.

These parameters are interrelated, and optimizing one can lead to trade-offs in the others.
For instance, reducing storage overhead with larger values of $N$ and $K$ increases the decoding time.
Therefore, instead of choosing solutions that excel in only one parameter, we retain only those on the Pareto front (Line~\ref{line:pareto}) that we call candidate mappings.
In Line~\ref{line:most_progress}, we compute for all candidates the progress that they made on each parameter relative to the other candidates.
Thus, if all candidates have a similar or equal storage overhead, they will not make much progress relative to others.
This favors solutions that are radically better than others on one or more parameters.

Finally, a score is calculated that combines the progress of the three parameters.
In order to account for the available capacity of the nodes, we compute the total \textit{system saturation} (see line~\ref{line:system_saturation}) using the same exponential function as before.
However, instead of evaluating individual nodes, we compute the overall free space across all storage nodes. 
This \textit{system saturation} is used to diminish the importance given to the duration parameter. 
As a result, this penalizes solutions that try to use parameters that are costly in terms of storage when the system is already saturated.
The candidate solution with the highest score is selected.

\begin{algorithm*}
    \caption{\drex{} System Capacity (\drex{} SC)}\label{algo:drex-sc}
    \begin{algorithmic}[1]
        \item \textbf{Task}: Determine for supplied $d$ the $\mathcal{M}$, $K$, $P$ that provide
    best trade-off between amount of data stored and throughput 
        \STATE Sort $\mathbb{S}$ in order of decreasing $F(S_i, t_{d})$
        \STATE $\mathbb{M}_{set} \gets$ First $2^{10}$ combinations of nodes\label{line:mapping}
        \FOR{each mapping $\mathcal{M} \in \mathbb{M}_{set}$}
            \STATE Choose $(K,P)$ such that $\lceil\frac{size(d)}{K}\rceil \times (K+P)$ is minimized and $Pr_{\textrm{avail}}(\mathcal{M},P,\Delta t) \geq RT(d)$\label{line:chooseK}
                \STATE $C \gets (K, P, \mathcal{M})$
                \STATE $duration(C) \gets min(B_w \in \mathcal{M}) \times \lceil\frac{size(d)}{K}\rceil +
                min(B_r \in \mathcal{M}) \times \lceil\frac{size(d)}{K}\rceil$ \hfill \textit{// Compute duration for given $d$, $\mathcal{M}$, $K$} \label{line:duration}
                \\\hspace{71mm} $+ T_{\textrm{encode}}(|\mathcal{M}|,K,size(d)) + T_{\textrm{decode}}(K,size(D))$\label{line:progress_time}
                \STATE $storage(C) \gets \lceil\frac{size(d)}{K}\rceil \times |\mathcal{M}|$  \hfill \textit{// Compute storage for given $d$, $\mathcal{M}$, $K$} \label{line:progress_storage}
                \STATE $saturation(C) \gets \sum_{k=1}^{|\mathcal{M}|} f(size(S_k) - F(S_k, t) + \lceil\frac{size(d)}{K}\rceil)$ \hfill \textit{// Compute saturation for given $d$, $\mathcal{M}$, $K$} \label{line:progress_saturation}
                \STATE Add $C$ to $candidate\_mappings$
         \ENDFOR
        \STATE $System\_Saturation \gets f(size(\mathbb{S}) - F(\mathbb{S}))$ for $f(x) = \exp(min(size(d) \in \mathbb{D} \texttt{ at time $t$}), 1/L)(size(\mathbb{S}),1)$
        \label{line:system_saturation}
        \STATE $\mathbb{C} \gets$ Pareto front of all $C \in candidate\_mappings$ based on \( \{duration(C), storage(C), saturation(C)\} \)
        \label{line:pareto}
        \STATE $\big(\textrm{duration}_{min},~ \textrm{storage}_{min}, ~\textrm{saturation}_{min}\big) \gets \big(\min_{c \in \mathbb{C}}{duration(c)}, ~\min_{c \in \mathbb{C}}{storage(c)}, ~\min_{c \in \mathbb{C}}{saturation(c)}\big)$
        \STATE $\big(\textrm{duration}_{max}, ~\textrm{storage}_{max}, ~\textrm{saturation}_{max}\big) \gets \big(\max_{c \in \mathbb{C}}{duration(c)}, ~\max_{c \in \mathbb{C}}{storage(c)}, ~\max_{c \in \mathbb{C}}{saturation(c)}\big)$
        \FOR{each candidate mapping \( C_i \in \mathbb{C}\)}\label{line:eachcandidate}  
                \STATE $\big(duration_{\textrm{progress}}(C_i), ~storage_{\textrm{progress}}(C_i), ~saturation_{\textrm{progress}}(C_i)\big) \gets \big(1 - \frac{duration(C_i) - \textrm{duration}_{min}}{\textrm{duration}_{max} - \textrm{duration}_{min}}, ~1 - \frac{storage_(C_i) - \textrm{storage}_{min}}{\textrm{storage}_{max} - \textrm{storage}_{min}}, ~1 - \frac{saturation(C_i) - \textrm{saturation}_{min}}{\textrm{saturation}_{max} - \textrm{saturation}_{min}}\big)$\label{line:most_progress}
            \STATE $Score(C_i) \gets (1 - System\_Saturation) \times duration_{\textrm{progress}}(C_i) + (storage_{\textrm{progress}}(C_i) + saturation_{\textrm{progress}}(C_i))/2$
        \ENDFOR
        \STATE Select candidate mapping $C_{best} = (K_{best}, P_{best}, \mathcal{M}_{best})$ with highest score
        \STATE Partition $d$ into $K_{best}$ chunks of size \( size(d)/K_{best} \), add $P_{best}$ parity chunks and distribute across nodes in \( \mathcal{M}_{best} \) 
    \end{algorithmic}
\end{algorithm*}

\begin{table}
        \caption{Measured scheduling overhead per data item (in milliseconds) for varying numbers of nodes.}
        \label{tab:complexity}
        \begin{center}
                \begin{tabular}{lrrrr}
                        \toprule
                        \textbf{Alg.} \(\downarrow\) \textbf{\# of nodes} \(\rightarrow\) & 10 & 50 & 100 & 500 \\
                        \midrule
                        GreedyMinStorage & 0.007 & 0.146 & 0.943 & 9.464  \\
                        GreedyLeastUsed & 0.001 & 0.003 & 0.006 & 0.027  \\
                        \drex{} LB & 0.001 & 0.015 & 0.057 & 2.462  \\
                        \drex{} SC & 0.488 & 2.871 & 6.671 & 343.3 \\
                        \bottomrule
                \end{tabular}
        \end{center}
\end{table}

Choosing $K$ (Line \ref{line:chooseK}) requires iterating over $L-1$ possible values of $K$ and computing the CDF for the Poisson-Binomial distribution. This results in a complexity of $O(L^{3})$ for this step.
This step is repeated for each mapping, which is at most $2^{10}$ times. This does not affect the asymptotic complexity, but it does affect the scheduling overhead for smaller values of $L$.
Adding the loop Line \ref{line:eachcandidate}, the overall complexity is $O(max(L^{3}, |\mathbb{C}|))$.
However, the scheduling overhead per data item remains reasonable even as the number of nodes scales, as shown in \autoref{tab:complexity}.

\section{Experimental evaluation}\label{sec:expes}

We evaluate our algorithms using a simulator written in C, since processing hundreds of terabytes of data in real-world experiments is costly. The simulator also allows us to easily explore different scenarios and node failures. 
We obtain data workloads (data items, sizes, arrival times) from four real-world datasets. The simulator processes data items using their release date from the dataset. It calculates transfer times using user-reported bandwidths without interference, i.e. in a best-case scenario. We implemented all strategies and state-of-the-art (SOTA) baselines in this simulator.\footnote{Our simulator, algorithm implementations, and details of storage nodes and datasets are available for reproducibility: \url{https://github.com/Double-Blind-975/Drex-repro-anonymous.git}.}
We calibrated the simulator with real measurements, and as will be shown by \autoref{fig:nodes_evolution_throughput} and \autoref{fig:dynres}, the results measured in the simulation are confirmed by real experiments.
We first describe the datasets and storage nodes used in the experiments and then evaluate our algorithms across different sets of nodes, datasets, and node failure scenarios.

\subsection{Datasets}\label{sec:datasets}

\begin{table}
\caption{Number of data items and per-item statistics for the four datasets used in our experiments.}
\centering
\label{tab:datasets}
\scalebox{0.77}{
{\setlength{\tabcolsep}{5pt}
\begin{tabular}{lrrrrr}
\toprule
 & & \multicolumn{4}{c}{\textbf{Per-item statistics}} \\ \cmidrule{3-6} 
\multicolumn{1}{c}{\textbf{Dataset}} & \multicolumn{1}{c}{\textbf{\# of items (m)}} & \multicolumn{1}{c}{\textbf{Avg.}} & \multicolumn{1}{c}{\textbf{Min}} & \multicolumn{1}{c}{\textbf{Max}} & \multicolumn{1}{c}{\textbf{Std}} \\ \midrule
MEVA & 4157 & 117.1 MB & 1.4 MB & 856.1 MB & 68.1 MB \\
Sentinel-2 & 256,351 & 475.9 MB & 2.7 MB & 969.9 MB & 256.5 MB \\
SWIM & 5214 & 23.4 GB & 1.0 B & 5329.5 GB & 177.0 GB \\
IBM COS & 47,529 & 2.6 GB & 0.2 MB & 1345.8 GB & 18.9 GB \\ \bottomrule
\end{tabular}
}
}
\end{table}

We use four datasets, summarized in \autoref{tab:datasets}.
The \textbf{MEVA} dataset~\cite{Corona_2021_MEVA} contains $4,157$ five-minute video clips, totaling 346 hours, captured from multiple cameras. It is widely used for human activity recognition research~\cite{Dave_2022_GabriellaV2}.
The \textbf{Sentinel-2} dataset~\cite{drusch2012sentinel} provides multispectral Earth imagery.
Imagery data are emitted and processed nearly daily, with a stable size. We sampled a representative portion of the dataset for benchmarking.
The Statistical Workload Injector for MapReduce (\textbf{SWIM})~\cite{chen2011case} is a tool for evaluating HDFS performance with realistic workloads, widely used in big data systems research~\cite{lee2019improving}.
The IBM Cloud Object Storage (\textbf{IBM COS}) dataset~\cite{eytan2020its} captures public data such as PDF documents, media files, database backups, or disk images.
It has been used to study caching policies~\cite{talluri2024exde}.
For benchmarking consistency, we standardize the total size of all requests performed for each dataset to 122 TB,
trimming the last three datasets and repeating the MEVA dataset to meet this size.

\subsection{State-of-the-Art Algorithm Baselines}
\label{sec:sota}

Within the scope of the proposed reliability model and problem formulation, we compare our algorithms against two state-of-the-art baseline algorithms
that are widely used in modern file systems and cloud platforms: three static erasure coding (EC) approaches and another with a varied replication factor.

\subsubsection{Static Erasure Coding}\label{sec:ec}

\begin{algorithm}[t]
    \caption{Erasure-Coding}\label{algo:ec}
    \begin{algorithmic}[1]
        \STATE \textbf{Function} EC($K$, $P$, $d$, $t$)
            \STATE Sort $\mathbb{S}$ by descending value of $B_w$
            \STATE $N \gets K + P$
            \STATE Create $N$ chunks, each of size $\lceil\frac{size(d)}{K}\rceil$
            \STATE Store chunks on first $N$ nodes that satisfy $\forall S_i \in \mathcal{M}$, $\lceil\frac{size(d)}{K}\rceil \leq F(S_i, t)$ and $Pr_{\textrm{avail}}(\mathcal{M},P,\Delta t_{d}) \ge RT(d)$
        \STATE \textbf{End function}
    \end{algorithmic}
\end{algorithm}

Popular solutions like HDFS \cite{hdfsdocrs} and Gluster \cite{glusterfsredhat} use static erasure coding to store data items.
We implement the default configurations of HDFS and Gluster, utilizing 10 or fewer storage nodes, as this will be the standard number of nodes in our experimental evaluation.
For HDFS, there are two erasure coding configurations: three data and two parity blocks, referred to as $EC(3,2)$; and six data and three parity blocks, referred to as $EC(6,3)$.
Gluster uses an $EC(4,2)$ configuration.
We describe how we apply this erasure coding in \autoref{algo:ec}.
For fairness, we assume that HDFS and Gluster prioritize nodes with the highest bandwidth.

\subsubsection{Erasure Coding with Varied Replication Factors}\label{sec:daos}

DAOS provides users with a selection of predefined erasure coding configurations \cite{10.1145/3581576.3581577, daosdoc}, including $EC(8, 1)$, $EC(8, 2)$, $EC(4, 1)$, and $EC(4, 2)$. For scenarios requiring higher reliability, DAOS also supports full data replication with configurations of $2\times$, $4\times$, or $6\times$ full copies. To ensure a fair evaluation, we prioritize choosing the configuration that meets the predefined reliability target $Pr_{\textrm{avail}}(\mathcal{M}_{d},P_{d},\Delta t_{d}) \ge RT(d)$ with the lowest storage overhead, aligning their performance with our objectives.

\subsection{Storage Nodes from Backblaze}\label{sec:backblaze_nodes}

We use statistics from Backblaze \cite{backblaze} to identify commonly used storage HDDs and their failure rates (SSDs and HDDs have similar failure rates \cite{pinciroli2020lifedeathssdshdds}).
We construct four distinct storage node sets where each node contains a single drive and each set contains 10 nodes:
(i) \textbf{Most Used}: 10 most used HDDs from Backblaze, representing a realistic scenario with popular hardware;
(ii) \textbf{Most Unreliable}: 10 HDDs with the highest failure rates, modeling a worst-case pathological scenario;
(iii) \textbf{Most Reliable}: 10 HDDs with the fewest failures; and
(iv) \textbf{Homogeneous}: 10 identical HDDs, each representing the most used HDD model from Backblaze.

\begin{figure}
     \centering
     \begin{subfigure}[b]{0.24\textwidth}
         \centering
         \includegraphics[width=\textwidth]{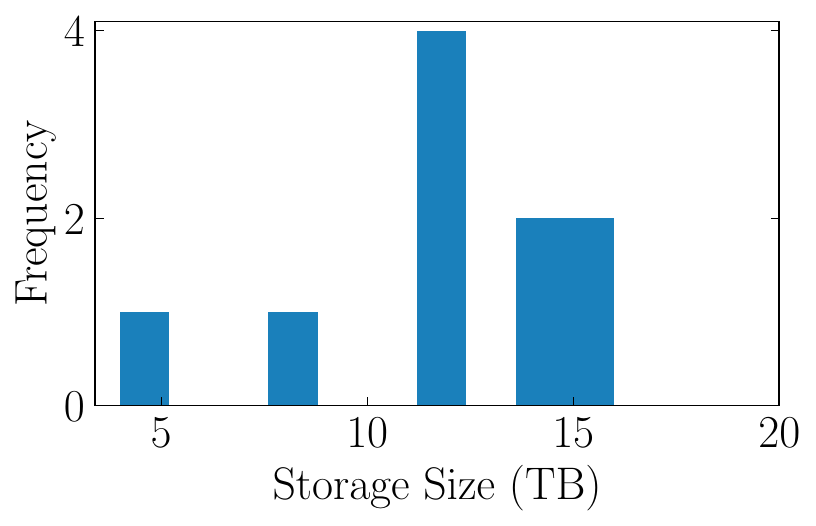}
         \label{fig:histogram-size-10mostused_TB}
     \end{subfigure}%
     \begin{subfigure}[b]{0.24\textwidth}
         \centering
         \includegraphics[width=\textwidth]{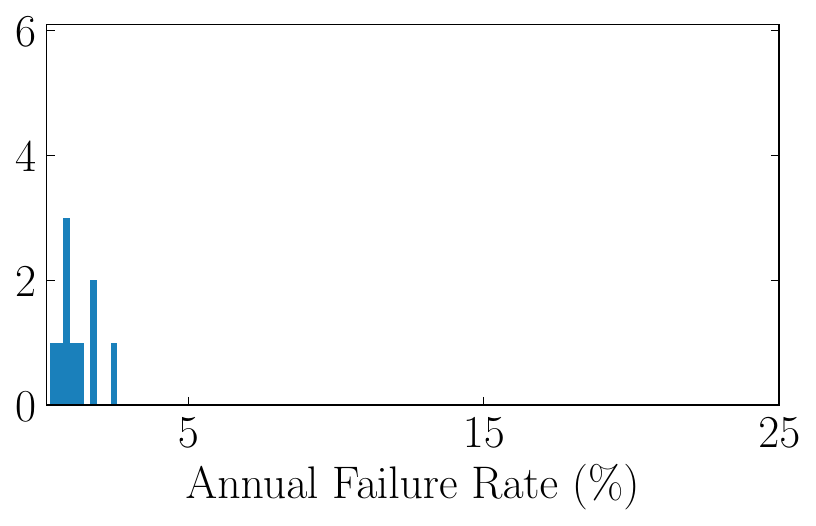}
         \label{fig:histogram-failurerate-10mostused_failurerate}
     \end{subfigure}
     \\[-0.3in]
    \begin{subfigure}[b]{0.24\textwidth}
         \centering
         \includegraphics[width=\textwidth]{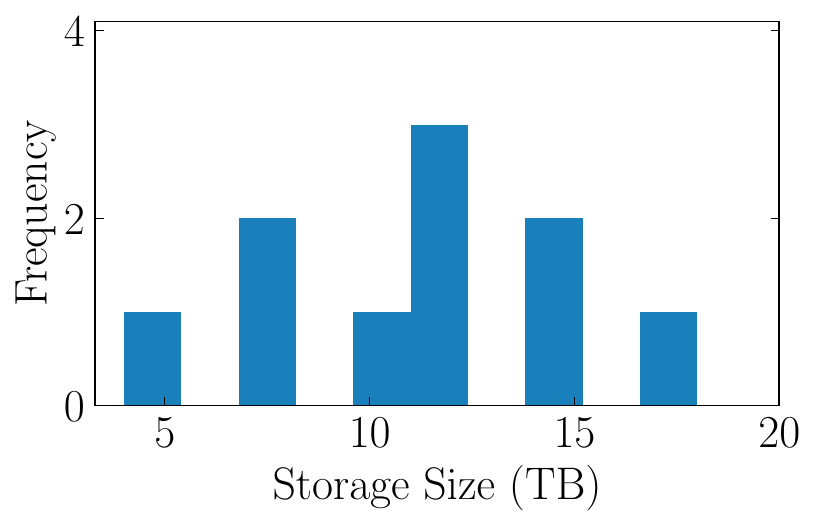}
         \label{fig:histogram-size-10mostused_storage_size}
     \end{subfigure}%
     \begin{subfigure}[b]{0.24\textwidth}
         \centering
         \includegraphics[width=\textwidth]{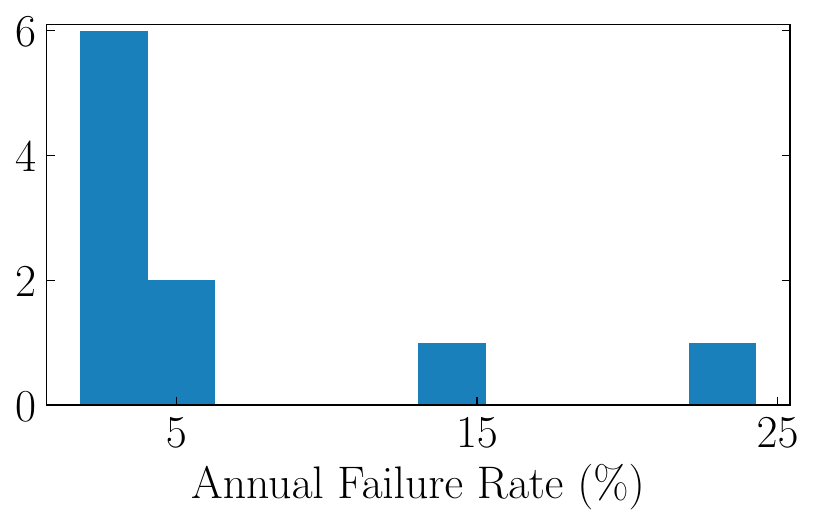}
         \label{fig:histogram-failurerate-10mostused}
     \end{subfigure}
        \caption{Distributions of storage size and annual failure rate for the \textit{Most Used} (top) and \textit{Most Unreliable} (bottom) sets.}
        \Description{Distributions of storage size and annual failure rate for the \textit{Most Used} (top) and \textit{Most Unreliable} (bottom) sets.}
        \label{fig:histogram-10mostused}
\end{figure}

\autoref{fig:histogram-10mostused} shows the distribution of storage size and annual failure rates for the \textit{Most Used} and \textit{Most Unreliable} sets. Storage sizes vary from 5 to 20 TB, and failure rates show significant variability for \textit{Most Unreliable}. 
Write bandwidths vary between 100 and 250 MB/s, and read bandwidths between 100 and 400 MB/s.
\autoref{tab:correlation-all_nodes_backblaze} shows the pairwise correlations across all HDDs.
The storage properties are largely independent, displaying no strong correlations, with the exception of read and write bandwidths.
Thus, it is unreasonable to simply rely on the most performant drives, which motivates exploiting the heterogeneity of the drives/nodes.
\begin{table}
\footnotesize
        \caption{Pairwise Pearson correlation among storage properties for all drives recorded by Backblaze.}
        \label{tab:correlation-all_nodes_backblaze}
        \begin{center}
                \begin{tabular}{lrrrr}
                        \toprule
                         & Storage Size & Write BW & Read BW & Failure Rate \\
                        \midrule
                        Storage Size & 1 & & & \\
                        Write BW & 0.614 & 1 & & \\
                        Read BW & 0.495 & 0.915 & 1 & \\
                        Failure Rate & 0.078 & -0.044 & -0.092 & 1 \\
                        \bottomrule
                \end{tabular}
        \end{center}
\end{table}

\subsection{Varying Reliability Target}\label{sec:results_most_used_nodes}

\begin{figure*}[t]
    \centering
    \includegraphics[width=1\linewidth]{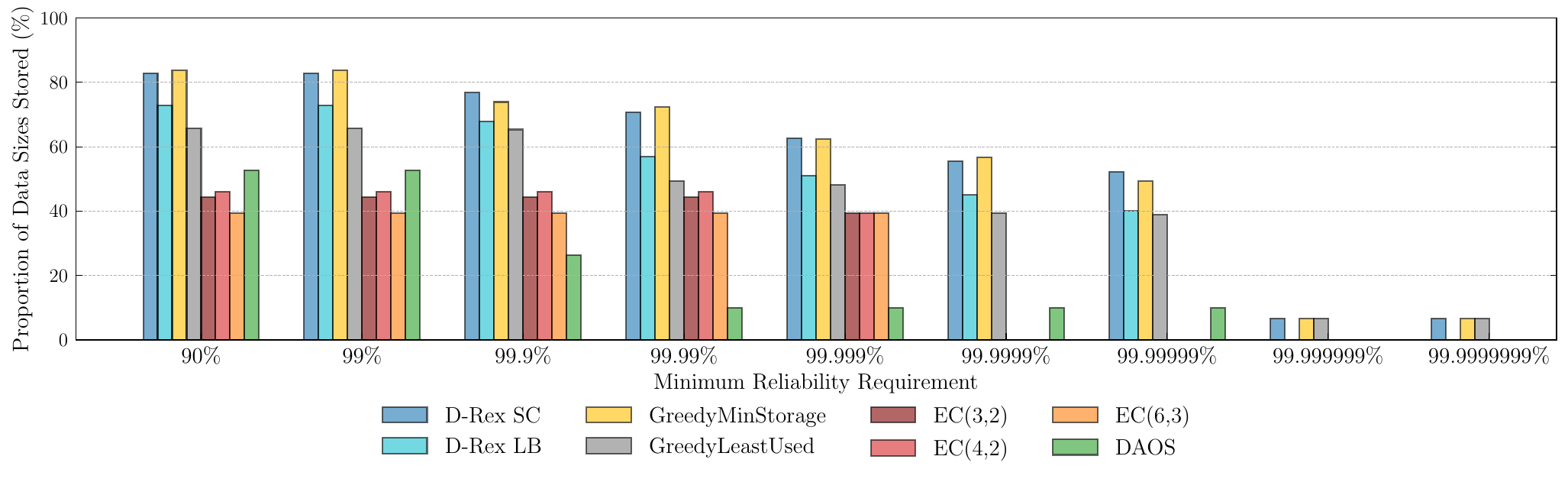}
    \caption{Evolution of proportion of data stored (the sum of the sizes of data stored by an algorithm relative to the sum of all data sizes in the workload) across varying reliability targets with the \textit{Most Used} nodes and MEVA dataset. Missing bars indicate that the required reliability level cannot be achieved.}
    \Description{Evolution of proportion of data stored (the sum of the sizes of data stored by an algorithm relative to the sum of all data sizes in the workload) across varying reliability targets with the \textit{Most Used} nodes and MEVA dataset. Missing bars indicate that the required reliability level cannot be achieved.}
    \label{fig:storage_and_efficiency_most_used}
\end{figure*}

\begin{figure}
    \centering
    \includegraphics[width=1\linewidth]{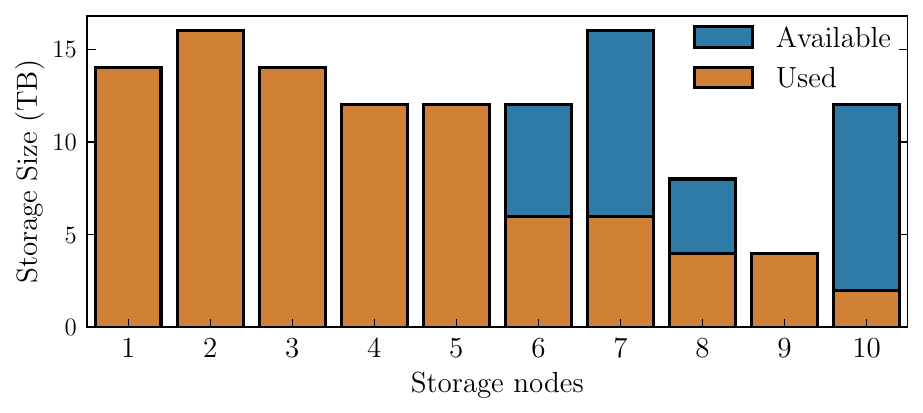}
    \caption{Consumed versus available storage for each node using EC(3,2) with a 90\% reliability target. 
     }
     \Description{Consumed versus available storage for each node using EC(3,2) with a 90\% reliability target.}
    \label{fig:distribution_hdfs32}    
\end{figure}

\autoref{fig:storage_and_efficiency_most_used} shows the proportion of data stored using the \textit{Most Used} nodes and MEVA dataset
while varying the target reliability.
At higher reliability targets, some bars are missing as some algorithms cannot deliver the required reliability level.
In all cases, \drex{} SC and GreedyMinStorage store the most data.
GreedyLeastUsed stores 10\% to 30\% less data items than \drex{} SC.
GreedyLeastUsed spreads the load evenly but always tries to create as few data chunks as possible. As a result, larger chunks of data are created to satisfy the reliability constraint, which increases the overall storage overhead of the stored data and thus allows fewer data items to be stored overall.
The SOTA algorithms (EC(3,2), EC(4,2), EC(6,4) and DAOS) perform poorly due to the fixed storage overhead associated with their fixed choice of $P$ and $K$.
Additionally, they do not load balance causing certain nodes to become saturated.
For example, \autoref{fig:distribution_hdfs32} shows the storage distribution of EC(3,2) with a 90\% reliability target.
The fastest nodes (1--5) are saturated; however, as EC(3,2) requires at least five nodes, no remaining items can be stored once node 9 is full, although significant storage remains available.
In contrast, \drex{} SC, \drex{} LB, and GreedyMinStorage avoid this limitation and fully utilize nodes.

\autoref{fig:storage_and_efficiency_most_used} shows that \textbf{\drex{} SC effectively incorporates the strengths of a greedy approach, storing at least 73\% more data than the SOTA strategies and as much data as a greedy algorithm that minimizes storage overhead}.

\subsection{Varying Storage Node Characteristics}

\begin{figure}
    \centering
    \includegraphics[width=1\linewidth]{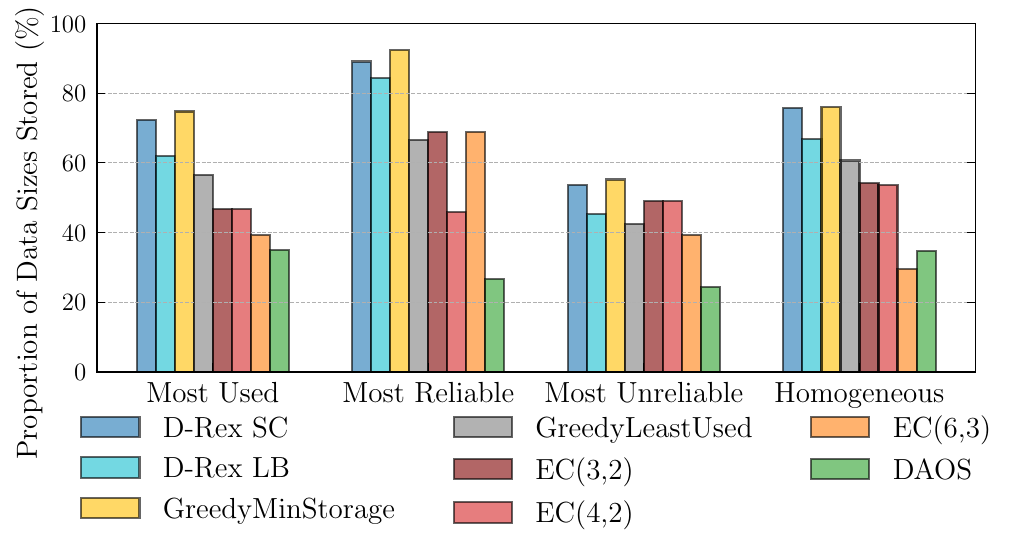}
    \caption{Proportion of data stored (the sum of the sizes of data stored by an algorithm relative to the sum of all data sizes in the workload) with different sets of nodes and random reliability targets between 90\% and 99.99999\% assigned to each item from the MEVA dataset.}
    \Description{Proportion of data stored (the sum of the sizes of data stored by an algorithm relative to the sum of all data sizes in the workload) with different sets of nodes and random reliability targets between 90\% and 99.99999\% assigned to each item from the MEVA dataset.}
    \label{fig:nodes_evolution}
\end{figure}

\begin{figure}[t]
    \centering
    \includegraphics[width=1\linewidth]{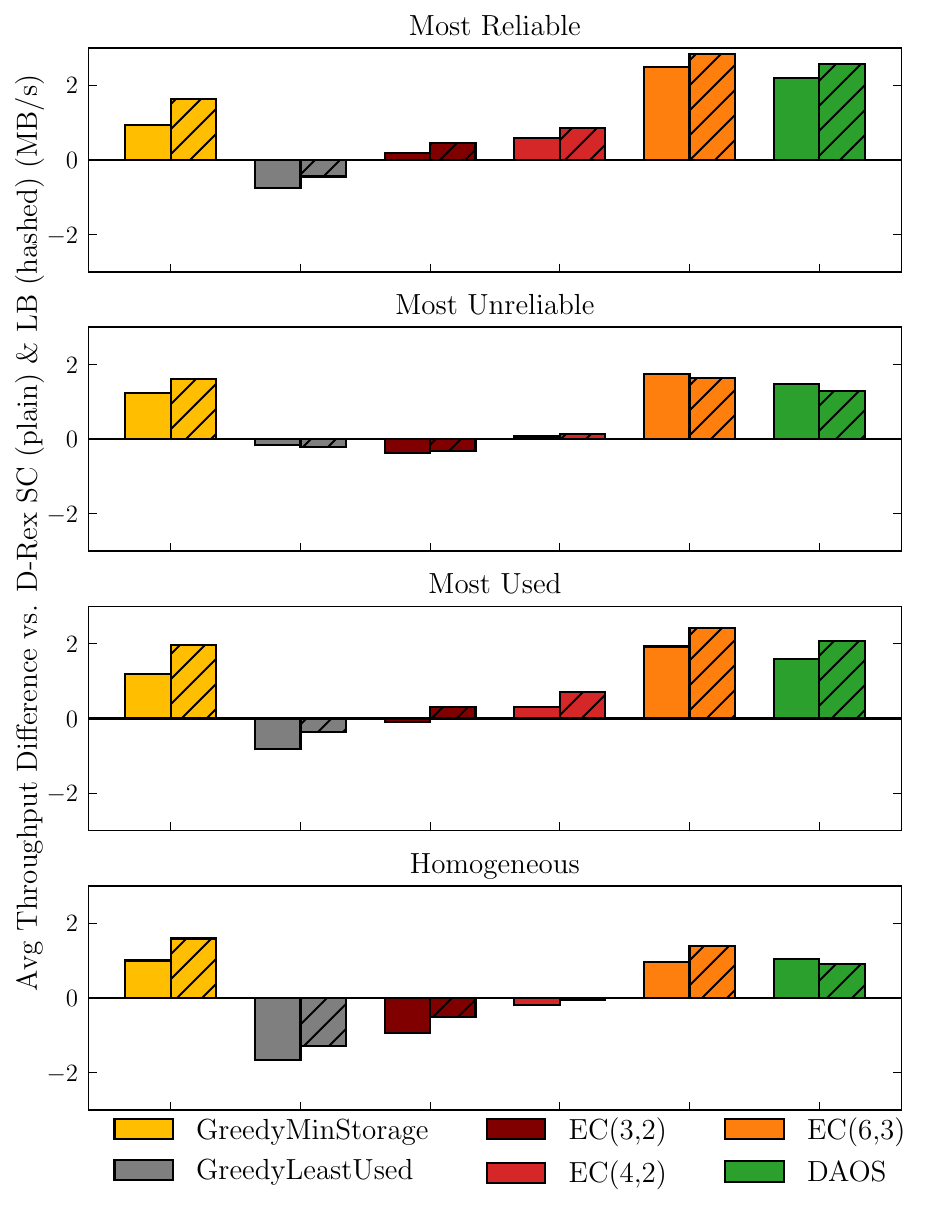}
    \caption{Average throughput difference when comparing \drex{} SC (plain) and \drex{} LB (hashed) against the rest of the algorithms using different sets of nodes and random reliability targets between 90\% and 99.99999\% assigned to each item from the MEVA. We compare equivalent data volumes by taking the minimum amount stored between the \drex{} and the compared algorithms. Values above 0 indicate that \drex{} algorithms are faster, while values below 0 indicate slower performance.}
    \Description{Average throughput difference when comparing \drex{} SC (plain) and \drex{} LB (hashed) against the rest of the algorithms using different sets of nodes and random reliability targets between 90\% and 99.99999\% assigned to each item from the MEVA.}
    \label{fig:nodes_evolution_throughput}
\end{figure}

\autoref{fig:nodes_evolution} shows the amount of data stored
on different node sets when each data item is assigned a reliability target (a ``number of nines'') chosen at random, as follows.
Let $x$ be a discrete random variable over the range 
$\{-1,0,1,2,3,4,5\}$. Define:
\[
f(x) = 
\begin{cases}
90 & \text{if } x = -1, \\
100 - 10^{-x} & \text{if } 0 \leq x < 5, \\
99.99999 & \text{if } x = 5.
\end{cases}
\]
Then, if $x \neq 5$, the reliability target is a random number chosen uniformly from the range $[\,f(x), f(x+1)\,]$, else it is 99.99999.

The storage proportion results are consistent with the previous results: both \drex{} SC and GreedyMinStorage store significantly more data, even when dealing with unreliable or homogeneous nodes, than the other algorithms.


\autoref{fig:nodes_evolution_throughput} compares the average data storage throughput of the \drex{} algorithms with the greedy and SOTA algorithms. Since different algorithms store varying amounts of data and use different subsets of nodes, a fair comparison requires matching the amount of data stored. For instance, a load-balancing strategy may utilize slower nodes, reducing throughput compared to a strategy that stores less data but uses only the same fast storage nodes. Thus, we compare the average throughput for each algorithm for the same data items.
For example, if EC(3,2) stores only the first 40\% of the workload, we compare its throughput with that of \drex{} SC/LB on the same set of data items.
We observe that the only algorithms faster than the \drex{} algorithms are GreedyLeastUsed, EC(3,2), and EC(4,2).
The largest difference occurs on the \textit{Homogeneous} node set, where algorithms using fewer nodes, such as GreedyLeastUsed and EC(3,2), benefit from reduced encoding/decoding time without the cost of not optimizing data placement for heterogeneity.
This result is to be expected, as \drex{} SC/LB were not designed for homogeneous scenarios.
In contrast, with diverse nodes like the \textit{Most Unreliable} or \textit{Most Used}, the difference with the \drex{} algorithms is negligible, approximately a 0.4 MB/s decrease at most.

\begin{figure}
    \centering
    \includegraphics[width=1\linewidth]{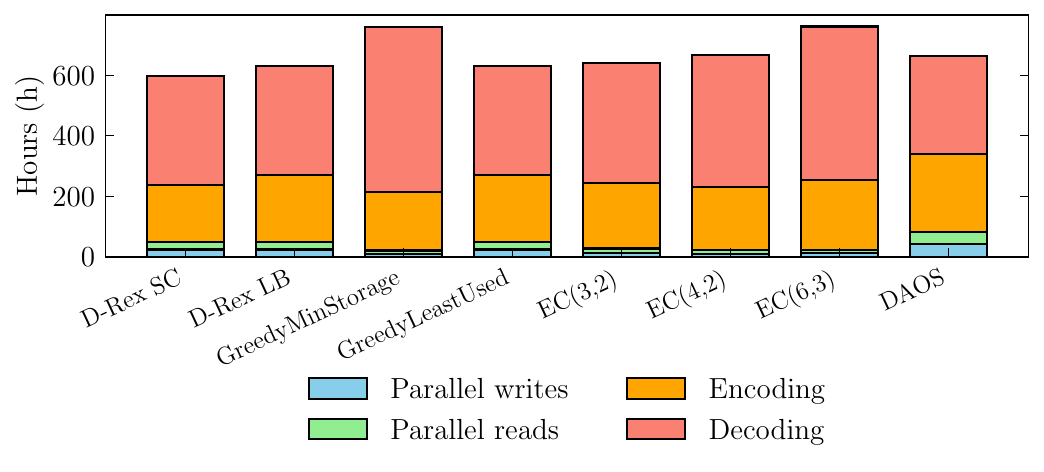}
    \caption{Time span of the data operations with a dataset that does not saturate the nodes and a reliability target of 99.99\%.}
    \Description{Time span of the data operations with a dataset that does not saturate the nodes and a reliability target of 99.99\%.}
    \label{fig:stacked}
\end{figure}

To better understand the performance of the different algorithms, we show in \autoref{fig:stacked} a breakdown of the time spent on each data operation for a subset of the MEVA workload.
(We use a subset so that all strategies can store all data items. Otherwise, a fair comparison would not be possible, as some strategies store more data and thus spend more time on data operations.)
We see that time spent reading \& writing is negligible compared to that spent encoding \& decoding because encode \& decode operations cannot be parallelized.
While GreedyMinStorage saves time on reading \& writing by creating smaller chunks, \drex{} LB and SC have faster encoding \& decoding because they balance storage and time overhead.

\textbf{
Overall, the \drex{} algorithms consistently store more data items across a heterogeneous set of nodes than SOTA algorithms while achieving comparable or better throughput (slow-down ranging from 0.2 to 0.8 MB/s for \drex{} SC and from 0.2 to 0.4 MB/s for D-Rex LB compared to the best SOTA result).
In a homogeneous scenario, although \drex{} stores more data items, it suffers greater loss in terms of throughput as our algorithms are not designed for such scenarios. 
In that case, GreedyLeastUsed is the better algorithm to use. Compared to a greedy algorithm that focuses on minimizing storage space, \drex{} SC stores as much data with significantly better throughput.
}

\subsection{Different Datasets}

\begin{figure}[t]
    \centering
    \includegraphics[width=1\linewidth]{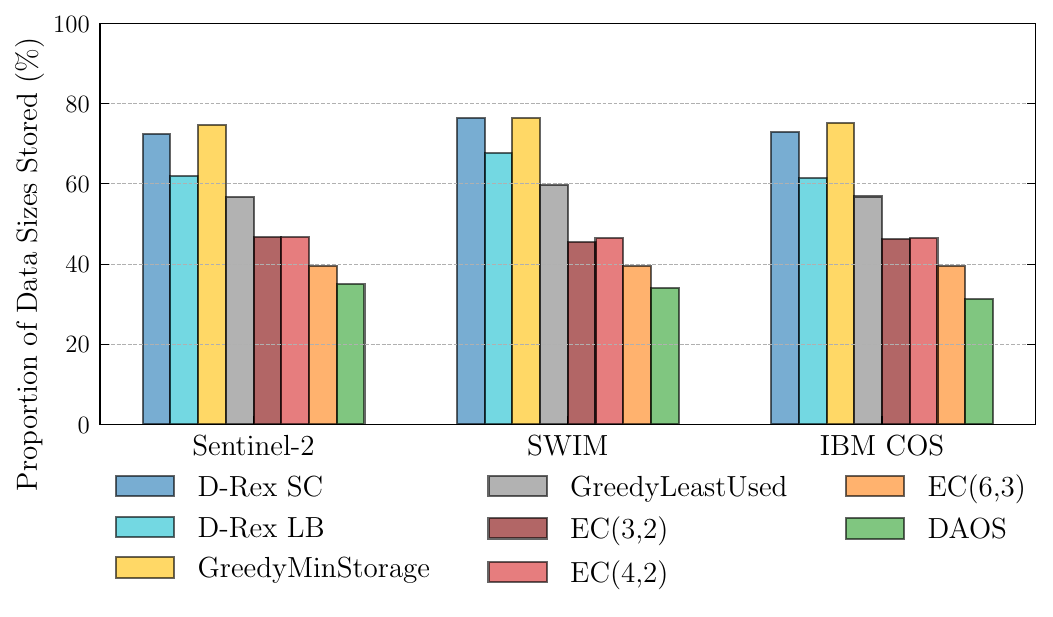}
    \caption{Proportion of data stored with the Sentinel-2, SWIM, and IBM COS datasets, a workload saturating all nodes, using the \textit{Most Used} nodes and random reliability between 90\% and 99.99999\% assigned to each item.}
    \Description{Proportion of data stored with the Sentinel-2, SWIM, and IBM COS datasets, a workload saturating all nodes, using the \textit{Most Used} nodes and random reliability between 90\% and 99.99999\% assigned to each item.}
    \label{fig:dataset_evolution}
\end{figure}

\begin{figure}[t]
    \centering
    \includegraphics[width=1\linewidth]{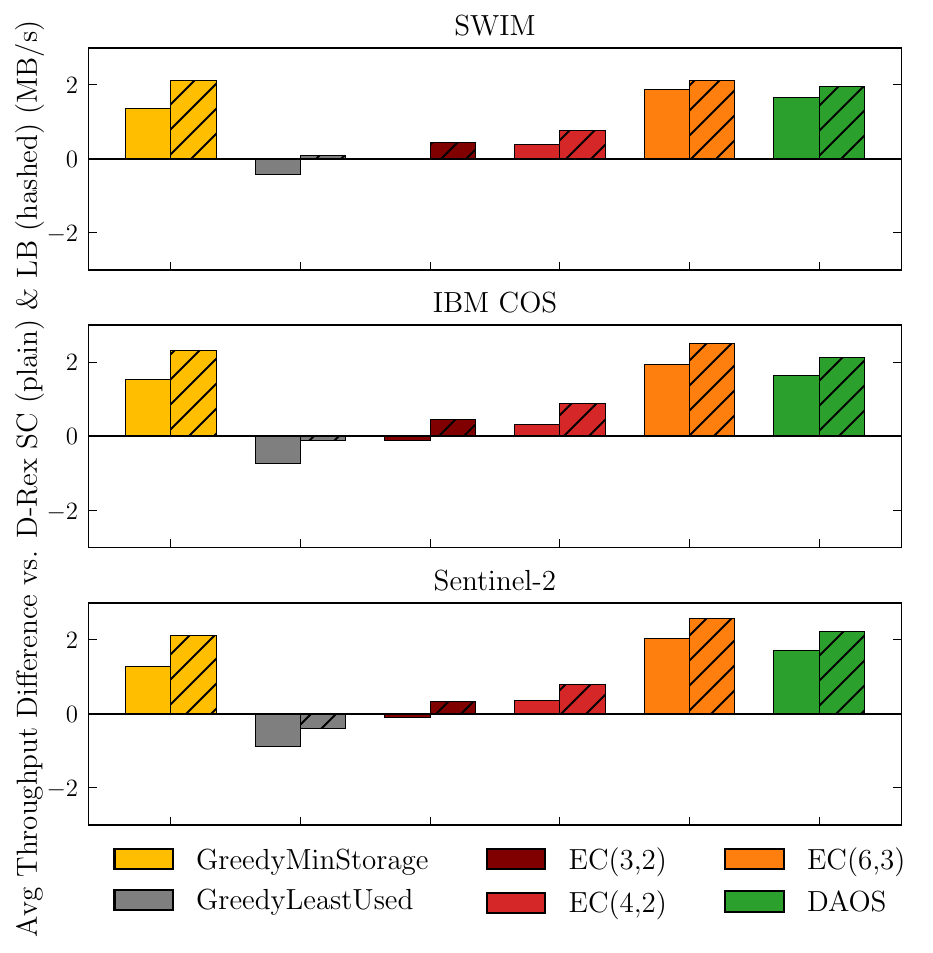}
    \caption{Average throughput difference when comparing \drex{} SC (plain) and \drex{} LB (hashed) against the rest of the algorithms using different datasets, the \textit{Most Used} set of nodes and random reliability targets between 90\% and 99.99999\% assigned to each data item.
    }
    \Description{Average throughput difference when comparing \drex{} SC (plain) and \drex{} LB (hashed) against the rest of the algorithms using different datasets, the \textit{Most Used} set of nodes and random reliability targets between 90\% and 99.99999\% assigned to each data item.}
    \label{fig:dataset_evolution_throughput}
\end{figure}

\autoref{fig:dataset_evolution} shows the amount of data stored using \drex{} and baseline algorithms for three different datasets: Sentinel-2, SWIM, and IBM COS (see \S\ref{sec:datasets}).  \autoref{fig:dataset_evolution_throughput} shows the difference in throughput between the \drex{} algorithms and the SOTA algorithms with the same datasets.
\textbf{Across several datasets with different data item counts and sizes, the \drex{} algorithms (SC and LB) store on average at least 45\% and 31\% more data than EC(3,2), EC(4,2), EC(6,3), or DAOS, respectively, while improving throughput or showing negligible degradation.
Compared to a greedy algorithm that optimizes for storage overhead, \drex{} SC stores almost as many data items while achieving higher throughput, by an average of 1.5 MB/s across the three workloads.
GreedyLeastUsed stores on average 21\% more data items while improving throughput compared to the SOTA.
}

\subsection{Resilience against Node Failures}\label{sec:expefailure}

\begin{figure}[ht]
    \centering
    \begin{subfigure}[b]{\linewidth}
        \centering
        \caption{Reliability target of $90\%$.}
        \label{tab:data_retained_failures_09}
        {\setlength{\tabcolsep}{5pt}
\scalebox{0.89}{
\begin{tabular}{lcccccc}
\toprule
\textbf{Alg.} \(\downarrow\) \textbf{\# failure} \(\rightarrow\) & \textbf{2} & \textbf{3} & \textbf{4} & \textbf{5} & \textbf{6} & \textbf{7}\\
\midrule
\drex{} SC & \colorcell{100} & \colorcell{100} & \colorcell{100} & \colorcell{56} & \colorcell{38} & \colorcell{28} \\
\drex{} LB & \colorcell{100} & \colorcell{100} & \colorcell{100} & \colorcell{54} & \colorcell{38} & \colorcell{23} \\
GreedyMinStorage & \colorcell{100} & \colorcell{100} & \colorcell{100} & \colorcell{52} & \colorcell{34} & \colorcell{28} \\
GreedyLeastUsed & \colorcell{100} & \colorcell{100} & \colorcell{100} & \colorcell{15} & \colorcell{0} & \colorcell{0} \\
EC(3,2) & \colorcell{100} & \colorcell{100} & \colorcell{100} & \colorcell{22} & \colorcell{0} & \colorcell{0} \\
EC(4,2) & \colorcell{100} & \colorcell{100} & \colorcell{82} & \colorcell{0} & \colorcell{0} & \colorcell{0} \\
EC(6,3) & \colorcell{0} & \colorcell{0} & \colorcell{0} & \colorcell{0} & \colorcell{0} & \colorcell{0} \\
DAOS & \colorcell{100} & \colorcell{100} & \colorcell{100} & \colorcell{0} & \colorcell{0} & \colorcell{0} \\
\bottomrule
\end{tabular}
}}
    \end{subfigure}

    \begin{subfigure}[b]{\linewidth}
        \centering
        \caption{Reliability target of $99.999\%$.}
        \label{tab:data_retained_failures_099999}
        {\setlength{\tabcolsep}{5pt}
        \scalebox{0.89}{

\begin{tabular}{lcccccc}
\toprule
\textbf{Algo} \(\downarrow\) \textbf{\# failure} \(\rightarrow\) & \textbf{2} & \textbf{3} & \textbf{4} & \textbf{5} & \textbf{6} & \textbf{7}\\
\midrule
\drex{} SC & \colorcell{100} & \colorcell{100} & \colorcell{100} & \colorcell{15} & \colorcell{0} & \colorcell{0} \\
\drex{} LB & \colorcell{100} & \colorcell{100} & \colorcell{100} & \colorcell{0} & \colorcell{0} & \colorcell{0} \\
GreedyMinStorage & \colorcell{100} & \colorcell{100} & \colorcell{100} & \colorcell{15} & \colorcell{0} & \colorcell{0} \\
GreedyLeastUsed & \colorcell{100} & \colorcell{100} & \colorcell{100} & \colorcell{15} & \colorcell{0} & \colorcell{0} \\
EC(3,2) & \colorcell{0} & \colorcell{0} & \colorcell{0} & \colorcell{0} & \colorcell{0} & \colorcell{0} \\
EC(4,2) & \colorcell{0} & \colorcell{0} & \colorcell{0} & \colorcell{0} & \colorcell{0} & \colorcell{0} \\
EC(6,3) & \colorcell{0} & \colorcell{0} & \colorcell{0} & \colorcell{0} & \colorcell{0} & \colorcell{0} \\
DAOS & \colorcell{85} & \colorcell{66} & \colorcell{41} & \colorcell{0} & \colorcell{0} & \colorcell{0} \\
\bottomrule
\end{tabular}
}}
    \end{subfigure}
    \caption{Proportion of data items retained after node failures under different reliability targets.}
    \Description{Proportion of data items retained after node failures under different reliability targets.}
    \label{fig:data_retained_failures}
\end{figure}

Tables~\ref{tab:data_retained_failures_09} and~\ref{tab:data_retained_failures_099999} present the results of an experiment where we simulate the failures of varying numbers of nodes.
The experiment uses the \textit{Most Unreliable} set of nodes and the MEVA dataset, which consists of 70 days of input data.
To simulate node failures, we calculate the daily failure rate of each node based on its known annual failure rate. Jobs are submitted as usual, and after one day of job submissions (tracked using the known jobs' submission times), we simulate node failures. For each node, we generate a random number between 0 and 1; if the random number is less than or equal to the node's daily failure rate, the node is considered to have failed.

The algorithms schedule data items as in previous experiments. When a node fails, they reschedule the lost chunks to meet the reliability target. If a data item cannot meet the target, its chunks are removed from all nodes, reducing the percentage of stored data shown in the tables. 
\autoref{tab:data_retained_failures_09} shows the amount of data retained after failures with a target reliability of 90\%.
The two \drex{} algorithms and GreedyMinStorage retain the most data by using more storage nodes to reduce storage overhead and larger values of $K$ and $P$ to survive more failures.
Overall, \drex{} SC retains the most data items.

When increasing the target reliability to 99.999\%, as shown in \autoref{tab:data_retained_failures_099999}, fewer failures can be survived across all algorithms, yet GreedyMinStorage and \drex{} SC still retain the most data items.
With a fixed $K$ and $P$, the static algorithms cannot achieve sufficient reliability after two nodes fail---maintaining 100\% reliability until that point.
\textbf{Node failures change the reliability that can be achieved with a given set of nodes. Therefore, dynamic strategies that can increase the number of parity chunks and the total number of nodes used are able to retain more data items after failures.}

\section{Validation on Real Infrastructure}\label{sec:real}

\begin{table}[t]
    \centering
    \caption{Infrastructure used to evaluate D-Rex on the Chameleon Cloud.}
    \label{tb:chainf}
    \scalebox{0.68}{
    \begin{tabular}{cccccc}
        \toprule
       \multirow{2}{*}{\textbf{Node(s)}} & \multirow{2}{*}{\textbf{Site}} & \multirow{2}{*}{\textbf{\# CPUs}} & \multirow{2}{*}{\textbf{RAM (GB)}} & \multicolumn{2}{c}{\bf Storage} \\ \cline{5-6}
          &  & &    & \textbf{Size (GB)} & \textbf{Drive}\\ \midrule
        0, 1 &	TACC &  	40 &	125 &	370 & INTEL SSDSC1BG40 \\
        2 &	TACC & 	40 &	126 &	2000 & Seagate ST2000NX0273 \\
        3 &	TACC & 	160 &	251 &	450 & Micron MTFDDAK480TDS \\
        4, 5 &	NRP & 48 &	125 &	200 & Seagate ST9250610NS \\
        6 &	UC & 	96 &	251 &	960 & Dell Express Flash CD5  \\
        7 &	UC & 	96 &	251 &	7600 & INTEL SSDPF2KX076TZ \\
        8 &	UC & 	96 &	187	& 240 & Dell MZ7KM240HMHQ0D3 \\
        9 &	UC & 	96 &	251 &	865 & INTEL SSDPF2KX076TZ \\
         \bottomrule
    \end{tabular}
    }
\end{table}

\begin{figure}
\includegraphics[width=\linewidth]{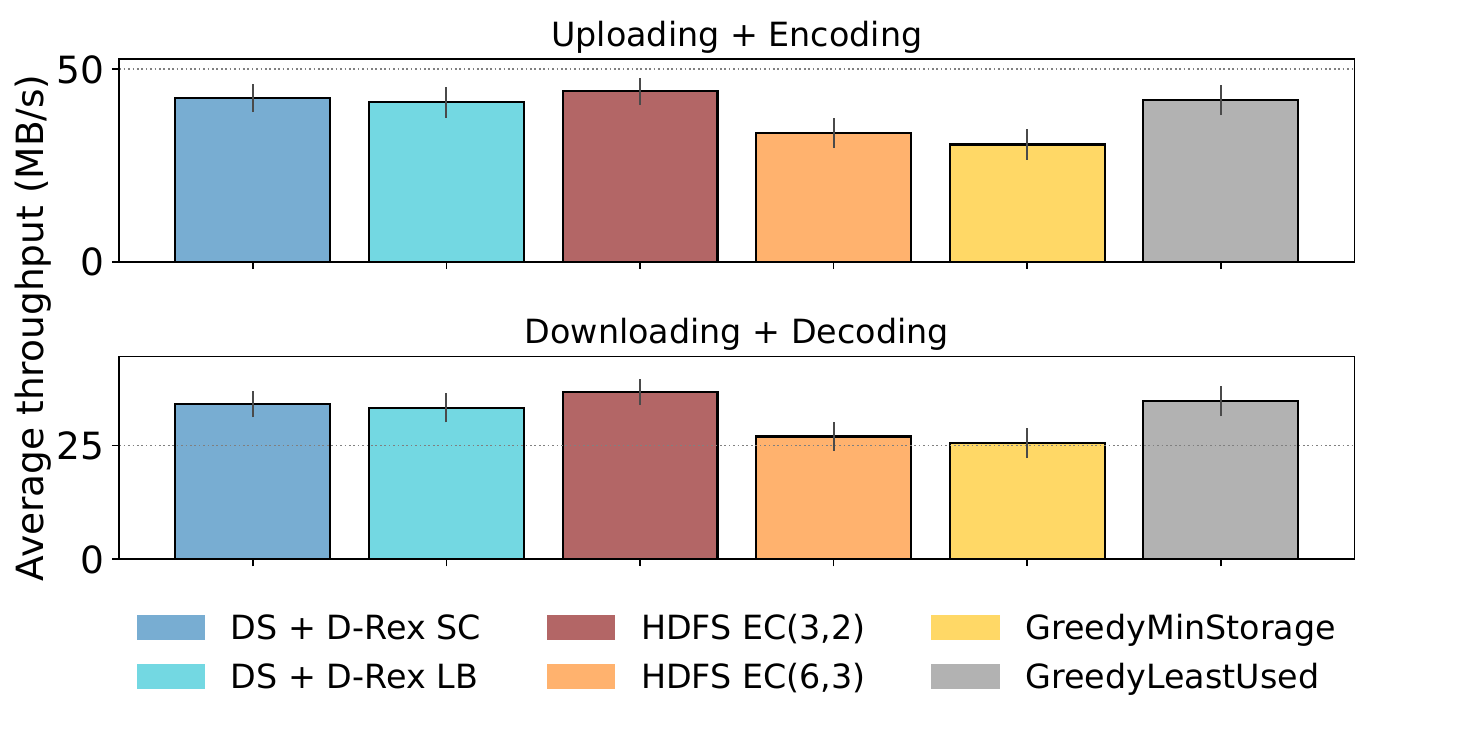}
\caption{Average throughput observed for uploading \& encoding and downloading \& decoding data in real infrastructure. DS means DynoStore. Averaged over 10 iterations.}
\Description{Average throughput observed for uploading \& encoding and downloading \& decoding data in real infrastructure. DS means DynoStore. Averaged over 10 iterations.}
\label{fig:dynres}
\end{figure}

To show the effectiveness of the \drex{} algorithms, we conduct real experiments with a wide-area storage system called DynoStore (DS) \cite{dynostore,sanchez2025dynostore}. DynoStore implements a federated storage overlay across distributed storage nodes using common interfaces called data containers. DynoStore can be configured to use different algorithms to map data items to storage nodes. Thus, while DynoStore manages the access to data and storage resources, we use \drex{} to control the chunking and placement of data items.
We deploy DynoStore with \drex{} using ten storage nodes distributed across three Chameleon Cloud sites: CHI@UC, CHI@TACC, and CHI@NRP~\cite{keahey2020lessons}. To achieve a heterogeneous setup, we selected nodes of various types and purposes. The hardware details are presented in Table \ref{tb:chainf}.

We compared the performance of \drex{} with HDFS.
For HDFS, we evaluated two erasure code policies using Reed-Solomon with parameters EC(3,2) and EC(6,3). \autoref{fig:dynres} shows the average throughput observed when uploading (top) and downloading (bottom) 1000 randomly generated data items with a resilience target of 99.999\%. The upload operations include the encoding of objects, whereas the download operations include the decoding of objects. The total size of the dataset is 113.5 GB with an average data item size of 116.3 MB.
\drex{} SC/LB, GreedyLeastUsed, and HDFS with EC(3,2) achieved similar performance: in \autoref{fig:dynres}, for data uploading, these algorithms yield a throughput of 42.4, 41.3, 41.9, and 44.2 MB/s, respectively.
HDFS with EC(3,2) performs slightly better because \drex{} utilizes a larger number of nodes, including some that are slower. Since the workload does not saturate the nodes, the advantages observed in simulations, where \drex{} achieved superior performance, are not yet apparent. When increasing the number of nodes used with HDFS EC(6,3), we observe that HDFS performs worse than D-Rex. In particular, \drex{} SC and \drex{} LB are 21\% and 19\% faster, respectively. A similar trend is observed in download operations.

\textbf{\drex{} stores more data items, retains more data items after failures (as demonstrated in the simulation results), and produces similar throughput to static algorithms from the literature (as demonstrated via experiments).}


\section{Conclusion}\label{sec:conclusion}

Distributed storage systems use erasure coding (EC) to increase reliability and reduce costs. However, EC is computationally intensive, and the varying capacity, performance, and failure rates of heterogeneous nodes present placement challenges. Existing solutions rarely take node heterogeneity into account when using EC. We address this gap by proposing two new dynamic schedulers that jointly select EC parameters and assign chunks to nodes: \drex{}-SC, which optimizes for overall system capacity, and \drex{}-LB, which balances load across nodes.
Through rigorous simulations, we showed that \drex{} is able to store on average 45\% more data items across different datasets and sets of nodes compared to classic state-of-the-art algorithms.
Through real-world experiments, we showed that \drex{} achieves comparable throughput compared to HDFS's erasure coding.
\drex{} SC provides the best overall performance compared to classic state-of-the-art algorithms in terms of both storage utilization and throughput, although its complexity increases with the number of nodes. In contrast, \drex{} LB offers a balanced tradeoff with reduced complexity. When storage is the primary concern, GreedyMinStorage is the best choice, while GreedyLeastUsed is superior when throughput is the sole focus.
In future work, we plan to enhance our heuristics by developing a system that selects the most appropriate algorithm based on hardware characteristics and user priorities.
Future work will also focus on exploring real-world deployments in storage systems such as Gluster.
By advancing the state-of-the-art in data storage algorithms and erasure coding, this work lays the foundation for resilient and high-performance storage leveraging distributed and heterogeneous storage nodes.

\begin{acks}
This material is based upon work supported by the U.S. Department of Energy (DOE), Office of Science, Office of Advanced Scientific Computing Research, under Contract DE-AC02-06CH11357. This publication is partially funded by NSF, under Grant CSSI-2411386 and by MICIU/AEI\-/\-10.13039/\-501100011033/``FEDER Una manera de hacer Europa'' under the project PID2022-138050NB-I00. We gratefully acknowledge the support of Chameleon Cloud for providing the computational resources (project CHI-231082) used in this research.
\end{acks}

\bibliographystyle{ACM-Reference-Format}
\bibliography{main}

\end{document}